\documentclass[fleqn,usenatbib]{mnras}

\usepackage{newtxtext,newtxmath}

\usepackage[T1]{fontenc}

\DeclareRobustCommand{\VAN}[3]{#2}
\let\VANthebibliography\thebibliography
\def\thebibliography{\DeclareRobustCommand{\VAN}[3]{##3}\VANthebibliography}

\usepackage{gensymb}    
\usepackage{graphicx}	
\usepackage{amsmath}	
\usepackage{xcolor, soul} 
\usepackage[normalem]{ulem} 
\usepackage{lscape}
\usepackage{rotating}
\usepackage{lineno}

\sethlcolor{yellow}

\newcommand{\cb}{^{\rm b}C_{\rm j}}
\newcommand{\cd}{^{\rm d}C_{\rm j}}
\newcommand{\ck}{^{\rm k}C_{\rm j}}

\newcommand{\etal}{et~al.}
\newcommand{\lya}{Ly$\alpha$}

\newcommand{\hi}{H\,{\scriptsize I}}
\newcommand{\his}{H\,{\scriptsize I}*}
\newcommand{\hii}{H\,{\scriptsize II}}
\newcommand{\hei}{He\,{\scriptsize I}}
\newcommand{\heis}{He\,{\scriptsize I}*}
\newcommand{\heist}{He\,{\scriptsize I}*\,$\lambda$3889}

\newcommand{\cii}{C\,{\scriptsize II}}
\newcommand{\ciis}{C\,{\scriptsize II}*}
\newcommand{\ciii}{C\,{\scriptsize III}}
\newcommand{\ciiist}{C\,{\scriptsize III}*}
\newcommand{\ciiis}{C\,{\scriptsize III}*\,$\lambda$1175}
\newcommand{\civ}{C\,{\scriptsize IV}}

\newcommand{\nv}{N\,{\scriptsize V}}
\newcommand{\oi}{O\,{\scriptsize I}}

\newcommand{\ovi}{O\,{\scriptsize VI}}
\newcommand{\mgii}{Mg\,{\scriptsize II}}
\newcommand{\alii}{Al\,{\scriptsize II}}
\newcommand{\aliii}{Al\,{\scriptsize III}}

\newcommand{\siii}{Si\,{\scriptsize II}}
\newcommand{\siiis}{Si\,{\scriptsize II}*}
\newcommand{\siiii}{Si\,{\scriptsize III}}
\newcommand{\siiv}{Si\,{\scriptsize IV}}
\newcommand{\pii}{P\,{\scriptsize II}}
\newcommand{\piii}{P\,{\scriptsize III}}
\newcommand{\suii}{S\,{\scriptsize II}}
\newcommand{\suiii}{S\,{\scriptsize III}}
\newcommand{\tiiii}{Ti\,{\scriptsize III}}
\newcommand{\crii}{Cr\,{\scriptsize II}}

\newcommand{\mnii}{Mn\,{\scriptsize II}}
\newcommand{\feii}{{\rm Fe\,{\scriptsize II}}}
\newcommand{\feiis}{Fe\,{\scriptsize II}*}
\newcommand{\feiii}{Fe\,{\scriptsize III}}
\newcommand{\coii}{Co\,{\scriptsize II}}
\newcommand{\niii}{Ni\,{\scriptsize II}}
\newcommand{\niiis}{Ni\,{\scriptsize II}*}
\newcommand{\znii}{Zn\,{\scriptsize II}}

\newcommand{\nhi}{\ensuremath{N_\mathrm{H\,\textsc{\lowercase{I}}}}}
\newcommand{\nh}{\ensuremath{N_\mathrm{H}}}
\newcommand{\dnh}{\ensuremath{n_\mathrm{H}}}

\newcommand{\kms}{\hbox{~km~s$^{-1}$}}
\newcommand{\cmcl}{\mathrm{cm}^{-2}}
\newcommand{\cmcb}{\mathrm{cm}^{-3}}
\newcommand{\A}{~\AA}
\newcommand{\ergs}{\hbox{~erg~s}^{-1}}
\newcommand{\rl}{$R_{\rm BLR}$--$L$}

\title[intrinsic damped Ly$\alpha$ absorber]{Discovery of a Damped Ly$\alpha$ Absorber in the Circumnuclear Zone of the FeLoBAL Quasar SDSS J083942.11+380526.3}

\author[Wu \etal]{
Shengmiao Wu,$^{1\ast}$\thanks{E-mail:wushengmiao@pric.org.cn}
Xiheng Shi,$^{1\ast}$\thanks{E-mail:shixiheng@pric.org.cn}
Nibedita Kalita,$^{2}$
Xiang Pan,$^{1}$
Qiguo Tian,$^{1}$ 
\newauthor
Tuo Ji,$^{1}$
Shaohua Zhang,$^{3}$
Xuejie Dai,$^{1}$
Peng Jiang,$^{1}$
Chenwei Yang$^{1}$
and Hongyan Zhou$^{1,4}$\thanks{E-mail:zhouhongyan@pric.org.cn}
\\
$^{1}$MNR Key Laboratory for Polar Science, Polar Research Institute of China, Shanghai 200136, China \\
$^{2}$Janusz Gil Institute of Astronomy, University of Zielona G{\'o}ra, ul. Szafrana 2, PL-65-516 Zielona G{\'o}ra, Poland \\
$^{3}$Shanghai Key Lab for Astrophysics, Shanghai Normal University, Shanghai 200234, China \\
$^{4}$Key Laboratory for Research in Galaxies and Cosmology of Chinese Academy of Science, Department of Astronomy, \\
  University of Science and Technology of China, Hefei 230026, China
}

\date{Accepted XXX. Received YYY; in original form ZZZ}

\pubyear{2024}

\begin{document}
\label{firstpage}
\pagerange{\pageref{firstpage}--\pageref{lastpage}}
\maketitle

\begin{abstract}
SDSS J083942.11+380526.3 ($z=2.315$) is a FeLoBAL quasar that exhibits visible Balmer absorption lines (H$\alpha$), implying a significant $n=2$ population. 
The quasar also shows an array of absorption lines, including \oi, \niii, \feii, \mgii, \aliii\, to \civ\ and \nv. 
The high-ionization absorption lines such as \civ\ and \siiv\ are revealed by slightly blueshifted BAL troughs. 
The resonance doublets such as \mgii\ and \aliii\ are saturated but did not reached zero intensity which indicates that the BLR is partially covered. 
Overall, however, the absorption is predominantly from low-ionization \feii\ lines, emitted from ground and excited states up to at least 3.814 eV. This implies that the absorbing gas spans the hydrogen ionization front and extends into the partially ionized zone where neutral hydrogen is certainly present. 
Notably, the hydrogen line spectrum of the quasar shows no signature of expected Ly$\alpha$ absorption. Instead, the line spectrum shows an unusual Ly$\alpha$ emission characterized by a fully filled emission line spectrum which is a composite of a strong narrow core superposed on a weak broad base.
Taking into account the effect of partial covering to BLR, we have extracted a strong DLA trough in Ly$\alpha$ emission region. 
To fit the spectrum, we performed photoionized model calculations and compared them to the observations. 
We found that photoionization modeling using CLOUDY can successfully reproduce the main characteristics of the quasar spectrum, 
and the predicted neutral hydrogen column density arising from the clouds responsible for the low-ionization absorption provides a good match to the extracted DLA trough. 
This indicates that both the DLA and the low-ionization absorption arise from the same medium that is roughly collocated with the dusty torus.
\end{abstract} 



\section{Introduction}\label{sec:intro}
\let\thefootnote\relax\footnote{
{\hspace*{-2mm}{$^\ast$}{ These authors contributed equally to this work}}}

Quasar absorption line systems are excellent probes of both the physical properties of the absorbing gas intrinsic to the quasars (intrinsic absorbers, which are physically related to the quasars) and the distribution of gaseous clouds in the universe (intervening absorbers or host galaxy gas, which are unrelated to the quasar). These systems are primarily described by their neutral hydrogen column density. 
  
Quasar spectra often show a plethora of absorption lines that are detached from or adjacent to the emission lines. These absorption lines are imprinted on the quasar spectra. Based on linewidths, the traditional broad absorption lines (BALs) are absorptions exhibited in the spectra of Quasars with velocity widths more than $2000 \kms$, and blueshifted by up to $\sim 0.2c$ relative to systemic redshift \citep{1981ARA&A.19.41W}.
BALs are generally understood to form in high-velocity outflows, possibly connected with accretion disk winds \citep{1995ApJ.451.498M,1995ApJ.455.448D,2010A&A.521A.57T}. However, \citet{2018ApJ.857.60A} found that at least $50\%$ of BALQSO outflows are at distances greater than 100 pc from the central source, and at least $12\%$ are at distances greater than 1 kpc. 
Based on the ions producing the absorptions, BALs are further classified as High-ionization BALs (HiBALs), low-ionization BALs (LoBALs), or iron low-ionization BALs (FeLoBALs) \citep{2006ApJS.165.1T}. The majority of BALs show absorption only from highly ionized species such as \civ\,, \nv\,, and \ovi\, (typically identified through the presence of \civ\ absorption) and are referred to as HiBALs. 
A minority of BALs, known as LoBALs, also exhibit absorption from lower ionized species such as \mgii, and \aliii. A small subset of LoBALs, called FeLoBALs, additionally show absorption from excited fine-structure levels or excited levels of \feii\ and/or \feiii\ \citep{1987ApJ.323.263H,2002ApJS.141.267H}. 
Additionally, the rarest absorption features in BALs are found in a handful of objects that exhibit absorption in their Balmer lines \citep{2002AJ.124.2543H,Aoki06,2007AJ.133.1271H,2015ApJ.815.113Z,2016ApJ.829.96S,2018ApJ.853.167S}.

Recently, in the first paper of a series of four, \citet{2022ApJ.937.74C} presented a detailed analysis of the outflows in 50 low-redshift ($0.66 < z < 1.63$) FeLoBAL quasars using SimBAL, a novel spectral-synthesis modeling software \citep{2018ApJ.866.7L}. 
The authors identified a group of eleven FeLoBAL quasars, a new class dubbed loitering outflow objects. These are characterized by low outflow velocities and high column density winds located close to the central engine, within the vicinity of the torus. 
The authors also found that FeLoBAL quasars with Balmer absorption are only found in BALs with log $R \lesssim 1$ [pc], suggesting that the presence of Balmer absorption lines can be used as an indicator for compact broad absorption line (BAL) winds (small log $R$ with high densities).

The term ``damped Lyman absorption" simply describes a phenomenon where the column density of neutral hydrogen is sufficient to show damping wings. Damped Ly$\alpha$ systems (DLAs) are historically defined as having a neutral hydrogen column density \nhi$ \geq 2 \times 10^{20}\, \cmcl$ \citep{1986ApJS.61.249W}. 
Depending on their origin with respect to the background sources, DLAs observed in the spectra of quasars can be divided into an entity either entirely separate from the quasar (intervening DLA) or loosely associated with the quasar (proximate DLA). 
Intervening DLAs are produced by neutral hydrogen gas located by chance along the line of sight to the background sources without being related to the sources themselves \citep{2019A&A.627A.32N}.
They were initially observed in QSO spectra and typically originate in the \hi\ regions of the high-redshift progenitors of present-day spiral galaxies \citep{1986ApJS.61.249W} or dwarf galaxies \citep{1988ApJ.329L.57T}. 

Associated DLAs may arise either in galaxies neighboring the quasar host or from \hi\ clouds within the quasar host galaxy itself. 
\citet{2013A&A.558A.111F} found a population of proximate DLAs (PDLAs) that do not fully cover the Ly$\alpha$ emission region of the background quasar \citep[see also][]{2016ApJ.821.1J,2018ApJ.858.32X}. 
In some extreme cases, a new class of PDLAs called ghostly DLAs as reported by \citet{2017MNRAS.466L.58F}, has been described in the literature \citep[see also][]{2020ApJ.888.85F,2021MNRAS.502.3855L}. In the case of ghostly DLAs, the BLR is not fully covered and the absorption system is only detected through its Ly$\beta$, Ly$\gamma$ and higher series lines as well as metal absorption lines (metal-selected). 
These ghostly DLAs are expected to be located closest to the central engine.

Additionally, associated DLAs are also common in long-duration gamma-ray burst (GRB) afterglow spectra, typically with much higher column densities at the burst redshift than it is observed in QSO-DLAs \citep{2004A&A.419.927V,2006ApJ.652.1011W}. The extremely large \hi\ column densities observed in many GRB-DLAs suggest an origin related to the host galaxy, presumably the GRBs' star-forming region.

DLAs are also characterized by the presence of multiple absorption lines from low ionization species such as \feii, \siii, \niii, and \znii, which are in their dominant ionization stages. 
The presence of the rare line transitions (e.g., \heis, Balmer series, excited fine-structure and other metastable, excited-state lines such as \ciis, \ciiist, \siiis, \niiis, \feiis) whether related to BAL systems or associated with DLA absorbers, can be used as various indicators or diagnostics, for example (1) \znii\, can be used as a metallicity indicator (\citealt{1998ApJ.498.256Z} and references therein), 
(2) \heis\ absorption lines are very useful diagnostics for the geometry and physical conditions of the absorbing gas in quasars \citep{2011ApJ.728.94L,2015ApJS.217.11L}, 
(3) the \ciiis\ multiplet lines or \siii\ and \siiis\ lines pairs can be valuable density diagnostic \citep{2005ApJ.631.741G,2019MNRAS.487.5041H}. Combining the density with photoionization modeling can be used to estimate the location of the absorbing gas, acting as a distance indicator, 
(4) FeLoBAL quasars with Balmer absorption can be used as an indicator for compact BAL winds \citep{2022ApJ.937.74C}, 
(5) in cases of partial covering, the observed depths of saturated lines are direct indicators of the covering fractions of optically-thick gas in that transition \citep{2016ApJ.819.99S,2019MNRAS.487.5041H}.

As is well known, absorption spectrum is an one-dimensional probe of the absorbers along our line of sight towards a quasar.
Conversely, emission line profiles encompass spatially integrated contributions, offering crucial spatial and kinematic insights into the structure of the line emitters.
To explore the spatial and kinematic structure of the DLAs, our initial step involves detecting the Ly$\alpha$ emission line.
A few rare examples of detecting the residual flux in the DLA trough have been reported in the literature.
However, the diverse origins of the flux within the DLA trough pose a challenge for their physical interpretation \citep{2009MNRAS.397.511B,2009ApJ.693L.49H,2014A&A.566A.24N}. 

In this paper, we report another ``loitering" outflow FeLoBAL quasar SDSS J083942.11+380526.3 (hereafter SDSS J0839+3805).
The UV spectrum of SDSS J0839+3805 includes an array of absorption lines, predominantly low-ionization \feii\ lines, emitted from ground and excited states up to at least 3.814 eV which appeared to be overlapping troughs. 
The quasar also exhibits saturated doublets absorption with partial coverage of the BLR.
The $K$- and $JH$-band spectra of SDSS J0839+3805 obtained with CISCO on the Subaru telescope also show Balmer and \heist\ absorption \citep{Aoki06}. 
The presence of a wide variety of absorption lines from a broad range of ions, especially rare features like Balmer series, \heis, excited fine-structure and excited-state lines, can place firm constraints on the physical properties and location of the outflowing gas. We conducted a comprehensive analysis of the spectrum of SDSS J0839+3805 and performed photoionized model calculations and compared them to the observations.

The paper is organized as follows. In Section \ref{sec:Obser}, we describe the observational data and analysis methods. An overview of the spectrum is presented in Section \ref{sec:Overview}. We developed a template fitting procedure that is described in Section \ref{sec:Method}. A uniform density model and its caveats are presented in Section \ref{sec:Uni_model}. 
In Section \ref{sec:Results}, we describe a turbulent, near-solar metallicity, two-component model and the results, while a more detailed information about atomic data of \feii\ for four datasets which is used in photoionization calculations is presented in Appendix~\ref{comparison}.
Discussion and summary are given in Section \ref{sec:Discussion}. Throughout the paper, we assume a cosmology consisting of $H_{0}=70\kms\mathrm{Mpc}^{-1}$, $\Omega_{\mathrm{M}}=0.3$ and $\Omega_{\mathrm{\Lambda}}=0.7$.

\begin{figure*}
  \centering
  \includegraphics[width=0.9\linewidth]{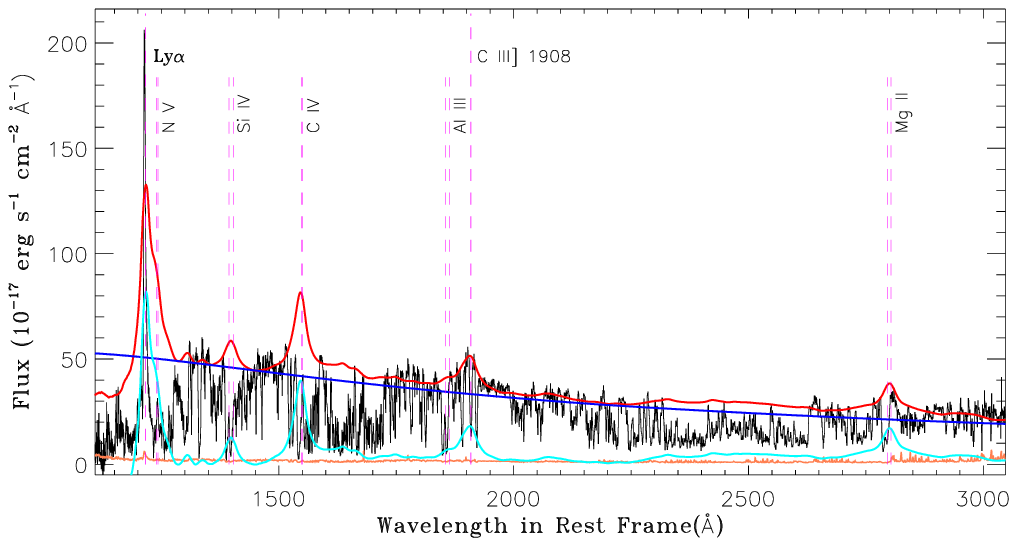}
  \caption{BOSS Spectrum of SDSS J0839+3805. We fitted EDR composite spectrum to the observed spectrum (in black) using an SMC extinction curve with flux at wavelength around 1324, 1350, 1480, 1960, 2238, 2645{\A} and longward 3020{\A}, that are regarded as continuum windows. The solid red/blue line represents the best fitted reddened EDR composite/continuum. The corresponding reddened emission lines are drawn in solid cyan line. The error spectrum (in coral) is also shown at the bottom.}
\label{SDSS J0839 SED}
\end{figure*}

\section{Observations and Data Reduction}
\label{sec:Obser}

The object was observed four times by the Sloan Digital Sky Survey (SDSS) from February 2002 to February 2016, and none of these spectroscopic observations show any significant change during this 14 years period (4.2 years in the quasar rest-frame). We selected the best spectrum from the Baryon Oscillation Spectroscopic Survey (BOSS) of SDSS-III from DR10 as our primary for scientific analysis which was observed on Modified Julian Date 55508 (calendar date:11/8/2010) with plate No.3765 and fiber No.392. The spectrum covers a wavelength range of 3650 $\sim$ 10400{\A}, and the resolution varies from $\sim$ 1500 at 3800{\A} to $\sim$ 2500 at 9000{\A}. 
 
We applied a Galactic extinction correction to the spectrum using the mean extinction curve from \citet{2007ApJ.663.320F} with the Galactic reddening value, $E(B - V)=0.037$ which was evaluated based on the extinction map traced by \citet{Schlegel98}. We then converted the observed wavelengths and shifted the observed spectrum to the quasar rest-frame by replacing the cosmological redshift by the systemic redshift (z = 2.315, given in SDSS DR7 catalog). The rest-frame spectrum covers the wavelength range of 1085 $\sim$ 3130{\A}.

\section{Continuum Fitting and Features Overview}\label{sec:Overview}

\subsection{Continuum Fitting with a Composite Spectrum}

We initiated the analysis by assessing the intrinsic reddening of the object in the quasar's rest frame, employing an SMC extinction curve \citep{Gordon03}. 
As a first approximation, we used the EDR composite spectrum \citep{Vanden01} as a representation of the unabsorbed quasar spectrum (i.e., the spectrum of the central source before interaction with the gas clouds), and fitted it to the observed spectrum. 
For simplicity, we omitted intergalactic extinction and assume that all dust reddening beyond the Milky Way are intrinsic to the quasar.  
Considering the ubiquitous strong UV absorption lines in the observed spectrum, we fitted a power law (reddened) to a set of emission- and absorption-free regions of the quasar spectrum  
and then we tried to match the dereddened observed spectrum with the EDR composite spectrum using the nonlinear least-squares procedure, MPFIT \citep{Markwardt09}, where we avoided the portions with strong line features to the greatest degree possible. 
If we consider the continuum points centered around 1324, 1350, 1480, 1960, 2238, 2645 {\A} , and longward of 3020 {\A} in the rest frame as continuum windows free of emission and absorption, the continuum points in both the composite spectrum and the dereddened SDSS J0839+3805 spectrum align with a power law. Following correction for intrinsic SMC-like reddening ($E(B - V)=0.03$), the best-fit power-law spectral index is determined to be $\alpha_{\lambda} = -1.46$.
The best-fitting EDR composite and the power-law continuum (reddened) are plotted in Fig.\ref{SDSS J0839 SED}.

\begin{table*}
  \centering  
  \caption{Measured line Parameters Based On Double Gaussians Fitting.}  
  \label{tbl-1} 
  \resizebox{2.\columnwidth}{!}{   
  \begin{tabular}{cccc}  
	 \hline\hline   
     Line & absorption-line center & FWHM &  Ionic Column Density \\
     & (km s$^{-1}$) & (km s$^{-1}$) &  cm$^{-2}$ \\ 
     \hline
     \niii\,$\lambda 1455$ & $-245.2\pm 5.8$/$-598.3\pm 24.2$ & $225.3\pm 13.2$/$251.5\pm 60.4$ & $(2.24\pm 0.12)$e+15/$(4.98\pm 1.00)$e+14 \\
     \niiis\,$\lambda 2207$ & $-219.0\pm 5.8$/$-572.1\pm 20.7$ & $249.1\pm 14.2$/$174.9^\textbf{a}$ & $(2.67\pm 0.13)$e+14/$(3.72\pm 0.68)$e+13 \\
     \feiis\,$\lambda 1761$ & $-239.3\pm 8.0$/$-530.9\pm 18.4$ & $197.9\pm 19.1$/$174.9^\textbf{a}$ & $(4.20\pm 0.35)$e+14/$(1.45\pm 0.25)$e+14 \\
     \crii\,$\lambda 2056$ & $-253.3\pm 7.4$/$-567.9\pm 16.8$ & $222.4\pm 17.2$/$229.1\pm 41.4$ & $(3.44\pm 0.23)$e+14/$(1.40\pm 0.22)$e+14 \\ 
     \znii\,$\lambda 2026$ & $-230.6\pm 5.0$/$-624.\pm 17.9$ & $230.1\pm 11.6$/$247.3\pm 56.0$ & $(9.80\pm 0.44)$e+13/$(2.56\pm 0.44)$e+13 \\ 
     \hi\,$\lambda 1215$ & $-951.5\pm 9.9$/$-1493.8\pm 20.2$ & $405.8\pm 17.7$/$512.5\pm 42.8$ & $(3.10\pm 0.16)$e+14/$(2.08\pm 0.17)$e+14 \\ 
     \hi\,$\lambda 1215^\textbf{b}$ & $-953.3\pm 6.1$/$-1486.6\pm 13.8$ & $365.6\pm 10.6$/$487.5\pm 27.4$ & $(7.17\pm 0.26)$e+14/$(3.59\pm 0.19)$e+14 \\ 
     H$\alpha$ &...... &...... & (2$\sim$3.2)e+13$^\textbf{c}$ \\
     \hline 
    \multicolumn{4}{l}{Notes. \niiis\,$\lambda 2207$ and \feiis\,$\lambda 1761$ transitions from lower levels at 1.254, 2.029 eV; $^\textbf{a}$ Upper limit set by the spectral resolution (R$\sim$1800).}\\
    \multicolumn{4}{l}{$^\textbf{b}$ The case for saturated Ly$\alpha$ absorptionon due to partial coverage to BLR; $^\textbf{c}$ cited from \citet{Aoki06} --derived from the curve of growth.}
  \end{tabular}
}
\end{table*} 

\subsection{An Overview of the Spectrum}

The spectrum of SDSS J0839+3805 shows unusual \lya\ emission features. The \lya\ emission is represented by a composite of strong and intermediate velocity width cores superposed on a relatively weak broad base. As expected, the observed spectrum also shows numerous absorption lines from a broad range of ions, in particular, of C, N, O, Mg, Al, Si, P, Cr, Fe, Ni, and Zn, including \feiis, and \niiis\ from their excited states up to 3.814 eV and 1.859 eV, respectively. 
The high-ionization absorption lines, such as \civ\,$\lambda\lambda 1548, 1551$, and \siiv\,$\lambda\lambda 1394, 1403$ in the spectrum of quasar are appeared as BAL troughs near the center of the corresponding lines. These troughs span a similar velocity width of $\sim 2000\kms$ from about 500 to $-1500\kms$ with respect to the systemic velocity, and slightly shifted to the blueward of the corresponding emission lines.
We also note that the two discrete absorption troughs in the blue wing of \lya\ spectrum within BAL troughs of high-ionization lines are most likely due to \lya\ absorptions. But we cannot exclude the possibility that the absorptions originate from unidentified lines.
Also, there are several relatively narrower absorption lines from low-ionization species, such as \feii, \siii, \pii, \suii, \niii, \crii, up to \znii. Most of these lines are heavily blended together in the lower-resolution SDSS spectrum, partly due to the associated transitions being very close in wavelength, and partly due to unknown turbulent and/or velocity gradients broadening in the absorption gas.  

\begin{figure}
  \centering
  \includegraphics[width=\linewidth]{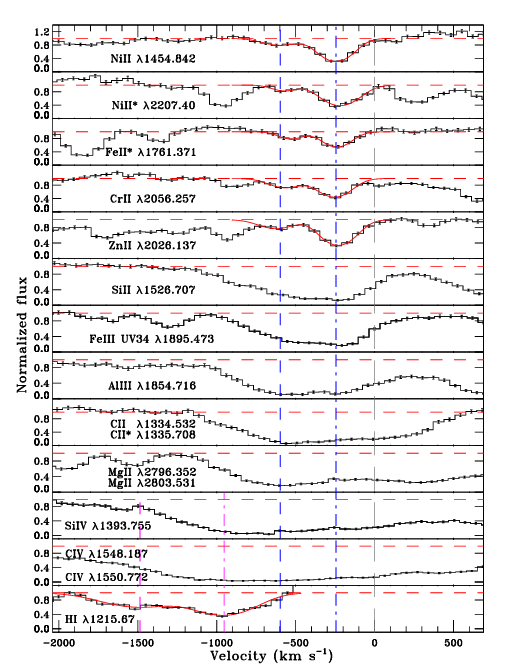}
  \caption{ Normalized flux for some typical absorption lines and Gaussian fits to six resolved absorption profiles in velocity space. The composite spectrum is simply normalized to unity shown by dashed red horizontal lines. The black lines with $\pm 1\sigma$ error bars show the observed data. The solid red lines are fitted with a double Gaussian for six resolved low/high-ionization absorptions. The dot-dashed/dashed blue (magenta) vertical line marks the centroid velocities of fitted Gaussian profiles for \niii\,$\lambda 1455$ (Ly$\alpha$) relative to the systemic redshift.}\label{abslps}
\end{figure}

Fig.\,\ref{abslps} shows the absorption profiles produced by ions, \niii, \feii, \crii, \znii, \siii, \feiii, \aliii, \cii, \mgii, \siiv, \civ, and \hi. Gaussian fits to five resolved absorption profiles of \niii\,$\lambda 1455$, \niiis\,$\lambda 2207$, \feiis\,$\lambda 1761$, \crii\,$\lambda 2056$, and \znii\,$\lambda 2026$ that have reasonable depth for low-ionization lines, as well as the resolved absorption profiles of Ly$\alpha$, are shown in solid red lines in both the first five and the last panels of the figure. The low-ionization lines show similar absorption profiles with two discrete velocity systems which are best fitted with a double Gaussian profile.
For example, the fitted Gaussian profiles to \niii\,$\lambda 1455$ give velocity shifts centered at $-245\pm 5.8\kms$ and $-598\pm 24\kms$ having full width at half-maximums (FWHM) of $225\pm 13\kms$ and $252\pm 60\kms$, respectively. The Ly$\alpha$ absorptions within BAL troughs of high-ionization lines can also be well fitted with a double Gaussian profile with a much higher blueshifted velocity. All these measurements are summarized in Table\,\ref{tbl-1}.
These two components of low-ionization lines are also present in strong absorption lines of both singly ionized species such as: \cii, \mgii, and \siii\ and doubly ionized species such as: \aliii, and \feiii\,UV 34. The ratio of absorption depths varies significantly between the two components. The high-ionization absorptions (e.g., \civ, and \siiv\,) exhibit BAL troughs that are much broader which fully encompass the velocity range of the low-ionization absorptions. 

From an immediate inspection of the quasar spectrum, it is obvious that the resonance absorption lines (e.g., \mgii, \aliii) are saturated but did not reach zero intensity (see Fig.\,\ref{abslps}), which indicates that the absorber has much higher column density and does not cover the background emission source completely. 
This high column density makes the usual much weaker lines from excited levels of \feii\ and \niii, and other less abundant iron group elements becoming detectable.     
This could greatly increase the number of narrow UV absorption lines from the low-ionization ions as expected for the observed spectrum of the quasar. 

It is important to note the numerous absorption lines of \feii\ which worth a brief description.
The observed \feii\ spectrum reveals approximately 80 \feii\ multiplets, with identifications extending up to at least UV 283 within the wavelength range redward of the Ly$\alpha$ emission line (see Fig.\,\ref{Absorber modelp}). The main \feii\ absorption features include: 
\begin{enumerate}
\item The profiles of \feii\ show that the depths of some strong absorption multiplets UV 1, UV 2, UV 3, and UV 62-63 near 2600{\A}, 2382{\A}, 2344{\A}, and 2750{\A} are somewhat shallower than that of the weak lines of UV 8 near 1635{\A}, and several other weak multiplets (e.g. UV 9, UV 38, UV 41, UV 44-45). 
\item Most of the transitions from 1550 to 1730 {\A} are due to \feii\,UV 38-46, UV 8, UV 68 and UV 84 multiplets, etc. Redward of this region, some other relatively resolved multiplets (not all) are UV 4-5, UV 60, UV 64-5, UV 78, UV 83, UV 145, UV 148, UV 161, UV 180, UV 191, UV 217, UV 234-235, UV 261, UV 263, UV 283, etc.
\end{enumerate} 
Considering the huge number of \feii\,lines, there must be some misidentifications.

\section{Template Fitting Method}\label{sec:Method}

Considering the heavy blending of the absorption features of SDSS J0839+3805, we strategized to generate a synthetic spectrum to fit the observed absorption line spectrum. 
For the template, we made the assumption that the shape of the function that describes the optical depth as a function of velocity (for any transition) is the same for all lines \citep{Korista92,Arav01}. In order to construct the synthetic model, we considered both the level population and the kinematic structure. 

\subsection{Multicomponent Fitting}\label{subsec:Multi-Covering}

Based on the ionization structure, a constant density slab of gas irradiated by an ionizing continuum of an AGN is divided into two distinct zones: (1) a highly ionized \hii\ zone (\hii-zone) at the illuminated face of the slab, where hydrogen is almost completely ionized (the hydrogen Str\"{o}mgren layer), and (2) an excited \hi\ zone (\his-zone) behind it, where a few to a few tens percent hydrogen is ionized \citep{Collin-S86}.  
Thus, almost all of the elements such as Fe, Ni and Zn are in singly ionized form in the partially ionized zone, whereas the doubly ionized species such as \aliii, are present in the Str\"{o}mgren layer but still get accumulated in considerable amounts inside the partially ionized zone \citep{deKool02}.   
 
The residual intensities measured at the lowest points (to avoid contamination from unknown emission to the utmost) in the saturated troughs of various transitions range from near zero (e.g., \cii\,$\lambda 1334$, \civ) to $\sim 0.2/0.4$ (e.g., \mgii) relative to full/broad emission. Also, a significant residual is present in the saturated low-ionization absorptions in strong broad emission line ranges (e.g., \siii\,$\lambda 1526$ in the blue wing of the \civ\ broad emission lines). Therefore, a more complicated multicomponent model for the background emission and stratified absorber must be taken into account. 

The background AGN emission is generally thought to be composed of a featureless continuum and a plethora of broad and narrow emission lines. 
However, the physically distinct sources (i.e., a continuum source and emission-line regions) have different sizes and geometries, thus, the covering factors can not be the same for all. Thus, in some cases, where the total background emission is only partially covered by the absorber, different viewing angle toward the distinct sources with different sizes would give different covering factors \citep{Barlow97,Gabel05}. 
Here, we model the background emission from the BLR and the compact accretion disk, separately. The BLR is further divided into two emitting regions of lower and higher ionization lines, which is consistent with a radially stratified ionization structure of BLR. 
This is supported by reverberation mapping studies which confirms that the emitting region of low ionization lines formed several times farther out than that of higher ionization lines \citep{1991ApJ.366.64C}. 
Thus, we applied separate covering factors, {$\cb^i$} and $\cd^i$ for the BLR and the disk, where the superscript i stands for the number of radially stratified layers in the absorbing gas along the line of sight, and the subscript j defines the different emitting regions of the source (e.g., the lower/higher ionization region of the BLR). 

As for example, if we consider three layers of the absorber where the covering factor decreases monotonically in the radial direction, then the observed residual flux can be given by the following equation:

\begin{eqnarray}
F_{j=l,h} =&&\sum_{k=b,d} F_k \Big[\ \ck^3\prod_{i=1}^3 e^{-\tau_i} + 1 - \ck^1 + \ck^1(\ck^1- \ck^2) e^{-\tau_1} \nonumber \\ &&+\ \ck^2(\ck^2-\ \ck^3) \prod_{i=1}^2 e^{-\tau_i}\Big],
\end{eqnarray}

where, $F_b$ and $F_d$ are the intrinsic broad emission-line and continuum fluxes, respectively. 

Dividing equation~(1) by $F_b + F_d$ gives 
\begin{eqnarray} 
I_{j=l,h} =&&\sum_{k=b,d} R_k \Big[\ \ck^3\prod_{i=1}^3 e^{-\tau_i} + 1 - \ck^1 + \ck^1(\ck^1- \ck^2) e^{-\tau_1} \nonumber \\ &&+\ \ck^2(\ck^2-\ \ck^3) \prod_{i=1}^2 e^{-\tau_i}\Big],  
\end{eqnarray}
where, $R_{b,d}=F_{b,d}/(F_b+F_d)$ is the normalized flux of the broad emission-line/continuum to the total intrinsic flux. These two equations can be easily extended to cases where the absorber is composed of more than three layers.

Using these equations with Cloudy, we developed a new algorithm that enables us to calculate the absorptions due to a continuous stratified medium, as well as a simplified compilation of atomic data that includes energy levels, wavelengths, and oscillator strengths of the transitions of selected ions from the National Institute of Standards and Technology Atomic Spectra Database (NIST)$^{1}$.
\footnote{$^{1}$The NIST Atomic Spectra Database was accessed at \\  https://www.nist.gov/pml/atomic-spectra-database.} 
In our compilation, we also added some observational results of line transitions that are available in the literature \citep{1991ApJS.77.119M,2003ApJS.149.205M,Fedchak99,2017ApJS.230.8C}.

For SDSS J0839+3805, it is reasonable to assume that the absorber fully covers the continuum source and partially covers the BLR. We also assume that both the \hii-zone and \his-zone have two sub-components, each having the same covering factor, $\cb^0=\ \cb^1$ and $\cb^2=\ \cb^3$. 
For an absorber composed of four layers with zero as initial value of i, the equations (1) and (2) can be simplified as
\begin{eqnarray}
F_{j=l,h}\ =&& F_d \prod_{i=0}^3 e^{-\tau_i} + F_b \Big[\ \cb^3\prod_{i=0}^3 e^{-\tau_i} + 1 -\ \cb^0 \nonumber \\ &&+\ \cb^1(\cb^1-\ \cb^2) \prod_{i=0}^1 e^{-\tau_i}\Big], 
\end{eqnarray}
\begin{eqnarray}
I_{j=l,h}\ =&& R_d \prod_{i=0}^3 e^{-\tau_i} + R_b \Big[\ \cb^3\prod_{i=0}^3 e^{-\tau_i} + 1 -\ \cb^0 \nonumber \\ &&+\ \cb^1(\cb^1-\ \cb^2) \prod_{i=0}^1 e^{-\tau_i}\Big], 
\end{eqnarray}
The factor is defined as fraction of the background light source(s) subtended by the absorbing cloud as seen by the observer. The observed depths of saturated lines are direct indicator of the covering fractions of optically thick clouds in that transition. Similarly, we can measure the uncovering fractions from the ratios of the measured residual intensities at the lowest points in the absorption troughs to the emission source. The covering factors used for the calculation are estimated by using the representative absorption lines of \cii\,$\lambda1334$, \siii\,$\lambda1526$, \civ\,$\lambda1548$, and \mgii\,$\lambda2796$ which are expected to be completely saturated. The measured covering factors are
$^{\rm d}C =1$, $^{\rm b}C_{h}^{0,1} =0.96$, $^{\rm b}C_{l}^{0,1} =0.83$, and $^{\rm b}C_{h}^{2,3} =0.74$, $^{\rm b}C_{l}^{2,3} =0.64$. 

\subsection{Optical Depth Template and Photoionization Simulations.}\label{sec:Simul} 

To construct the synthetic model, not only the kinematic structure but also the level populations were required. 
An essential ingredient required by the synthetic model is the optical depth template that describes the optical depth as a function of velocity extracted from the reference absorption line profile. 
This template is best derived from the observed profile of an unblended absorption line of intermediate depth that occurs in a part of the spectrum where the continuum is well defined, especially in the relatively smooth part of the spectrum free of broad emission lines. 
Once the optical depth template is known, the optical depth as a function of velocity for any transition is the same as that for the optical depth template with any arbitrary scale factor. The scale factors or ``multipliers'' for other ions are identical to the ratios of the ionic column densities of associated ions to that of the reference ion. 
The level populations of all ions that give rise to absorption lines in the spectrum can be calculated using the large-scale synthesis code Cloudy (version 17.02, \citet{2017RMxAA.53.385F}).

We choose \niii\,$\lambda 1455$ absorption from the ground level as our optical depth template for low-ionization absorption lines. This absorption has two discrete velocity systems that were fitted with a double Gaussian profile.   
We found obvious redshifted absorptions \citep[see also][]{2022ApJ.937.74C} in the wavelength ranges, \nv\,$\lambda 1242$ to \suii\,$\lambda 1250$ and \civ\,$\lambda 1550$ to \feii\,$\lambda 1559$ in the red wings of high-ionization absorption lines, \nv\ and \civ\ (see Fig.\,\ref{Absorber modelp}). Presumably these absorptions might be produced by highly ionized gas in an inflow (Wu et al. 2024, in preparation). 
Due to heavy saturation and low resolution, we did not fit the high-ionization absorption troughs with a theoretical model. Instead, we choose a Gaussian for an outflow besides a template absorption from Ly$\alpha$ at relatively high blueshifted velocity of $\sim 1000\kms$.
These were manually adjusted to get a suitable absorptions desired in the troughs of \civ\ and \nv.

The absorbing cloud was modeled as plane-parallel slab of gas exposed to ionizing radiation from the central engine, where the gas is assumed to have an uniform density and chemical composition. Initially, we applied solar abundance and zero turbulence, and the medium was assumed to be free of dust. The photoionization model was parameterized by ionization parameter $U$, number density of hydrogen, $\dnh$ and column density of hydrogen, $\nh$. A continuum similar to that deduced by \citet{Mathews87} was used as incident ionizing radiation (see also \citet{2013MNRAS.435.133H,2021ApJ.907.12S} for the influence from different SEDs). To reproduce the complex \feii\,absorption, we incorporated the full 371 levels $\mathrm{Fe}^+$ model in our simulation, including all levels up to 11.59 eV$^{2}$. 
To evaluate the accuracy of atomic data of \feii, we have introduced a comparison of atomic data for selected lines of \feii\ for four datasets from \citet{Verner99}, \citet{2019MNRAS.483..654S}, \citet{2018PhRvA..98a2706T} and CHIANTI$^{3}$. Some of these results from the latter three datasets were incorporated into the atomic data of \feii\ (see Appendix~\ref{comparison} for further details) to produce synthetic \feii\ spectra.

\let\thefootnote\relax\footnote{
\textsuperscript{2}{ 
It is worth pointing out that the atomic data for \feii\ in \citet{Verner99} seems not fully consistent with their 371 levels ranging in energy up to 93487.650 $\mathrm{cm}^{-1}$, or 11.59 eV. In their atomic data for \feii, Both level 114 and 228 with energies 58631.531 $\mathrm{cm}^{-1}$ and 77230.9 $\mathrm{cm}^{-1}$ have somehow been missed and shift upward by their next levels 58666.258 and 77742.73 $\mathrm{cm}^{-1}$ respectively. In reality, the lack of the last two levels were added by levels 93830.977 $\mathrm{cm}^{-1}$ and 93840.344 $\mathrm{cm}^{-1}$, constitute all 371 levels with energies below 11.635 eV.}}

\let\thefootnote\relax\footnote{
\textsuperscript{3}{ 
CHIANTI is an atomic database and software package which are available at https://chiantidatabase.org. }}

Initially, we generated a grid of photoionization models in the $U$-$\dnh$-$\nh$ space with the parameter ranges, $-4<\log\,U<4$, $0<\log\,\dnh(\cmcb)<15$, and $19<\log\,\nh (\cmcl)<24$ in steps of 0.2 dex for all of them. Among all the models in the initial grid of parameters, the gridpoints, which generate a synthetic spectrum with smallest $\chi^2$ compared to the observed spectrum were found. 
This allowed us to obtain a rough estimate of the parameter ranges. 
Then, a fine grid with halved increment was calculated by relaxing the assumptions of zero turbulence and solar abundance using the code Cloudy (version 17.02, \citet{2017RMxAA.53.385F}). This step helped us to narrow down the allowed values in the parameter space neighboring the found gridpoints to further constrain the absorber's properties.

\section{A Uniform Density Model and Its Caveats}\label{sec:Uni_model}

The observed spectrum of SDSS J0839+3805 shows numerous absorption lines which provides an unique opportunity to place firm constraint on the absorber's physical properties by simultaneously taking account of as many absorption lines as possible. The photoionization simulations for a small range of parameters can closely reproduce most of the observed spectrum features in the synthetic spectra.

\subsection{A Uniform, Homogeneous Absorber}\label{subsec:Uniform}

We performed a series of photoionization calculations using the code Cloudy and compared the output to the observations. In photoionization modeling, first we adopted a uniform density model, mainly focused on low-ionization absorption lines. The synthetic spectrum includes contributions from \hi, \hei, \cii, \ciii, \civ, \nv, \oi, \mgii, \alii, \aliii, \siii, \siiii, \siiv, \pii, \piii, \suii, \suiii, \tiiii, \crii, \mnii, \feii, \feiii, \coii, \niii\ and \znii. 
Ionic column densities for some typical lines and emission-line ratios of \civ/Ly$\alpha$ and \nv/Ly$\alpha$ predicted by photoionization simulations are listed in Table\,\ref{tbl-2}. The metallicity and turbulent velocity are considered as free parameters.
Among the models with $\log\,\dnh(\cmcb)=7.0$, the solar metallicity models can be ruled out by our data. The subsolar metallicity models yield acceptable results as a whole, except that the ionic column density of H$\alpha$/\znii\ is much higher/lower than the observed data. Model \#7 with $\log\,U=-1.2$, $\log\,\dnh(\cmcb)=6.6$, and $\log\,\nh(\cmcl)=22.3$ for which we adopted solar metallicity of $Z=Z_\odot$ and turbulent velocity, $v=50 \kms$ seems roughly in agreement with the observation. However, this model overpredicts the ionic column densities of ground and low excitation levels of \feii, and \niii, and thus, we referred it as ``preliminary'' solution for the uniform density model. The photoionization modeling yields a radial distance of $R_{\mathrm{abs}} \approx$ 32 pc from the nucleus for the absorbing clouds (see Section \ref{subsec:Location}). 

\begin{table*}
  \centering
  \caption{A Uniform, Homogeneous Absorber Model.} 
  \label{tbl-2}
  \resizebox{2.\columnwidth}{!}{
  \begin{tabular}{lllcccccccccccc} 
    \hline
    Model & Turbulence  & Metallicity & \multicolumn{7}{c}{Predicted Ionic Column Density} && \multicolumn{2}{l}{Emission-line Ratio} \\
    \cline{4-10}\cline{12-13}  
    $\log(U/n_{\rm H}/N_{\rm H})$ & (km s$^{-1}$) & (Z$_{\odot}$) & Ly$\alpha$ & H$\alpha$ & Zn\,{\tiny II} & Fe\,{\tiny II} & Fe\,{\tiny II}\,UV191 & Al\,{\tiny III} & Si\,{\tiny IV} && C\,{\tiny IV}/Ly$\alpha$  & N\,{\tiny V}/Ly$\alpha$\\
    \hline  
    \#1: -1.3/7.0/22.30 & 0 & 1 & 9.66e+21 & 8.69e+13 & 1.55e+14 & 6.09e+16 & 6.09e+13 & 3.60e+15 & 9.48e+16 && 0.26 & 0.008 \\
    \#2: -1.3/7.0/22.30 & 100 & 1 & 9.91e+21 & 4.52e+13 & 1.51e+14 & 5.88e+16 & 1.82e+14 & 3.74e+15 & 9.42e+16 && 0.26 &0.014 \\
    \#3: -1.3/7.0/22.00 &100 &1 &2.17e+21 &1.29e+13 &2.48e+13 &1.51e+16 &8.70e+13 &1.90e+15 &9.68e+16 && 0.27 &0.014 \\
    \#4: -1.15/7.0/22.35 &50 &0.3 &8.23e+21 &1.46e+14 &1.75e+13 &1.68e+16 &9.71e+13 &1.54e+15 &3.81e+16 && 0.30 &0.013 \\
    \#5: -1.15/7.0/22.35 &100 &0.3 &8.42e+21 &1.02e+14 &1.67e+13 &1.68e+16 &1.23e+14 &1.59e+15 &3.79e+16 & &0.30 &0.016 \\
    \#6: -1.20/6.6/22.30 &0 &1 &8.33e+21 &3.36e+13 &1.19e+14 &6.08e+16 &2.79e+13 &3.57e+15 &1.17e+17 && 0.25 &0.009 \\
    \#7: -1.20/6.6/22.30 &50 &1 &8.48e+21 &2.26e+13 &1.17e+14 &6.10e+16 &7.28e+13 &3.65e+15 &1.16e+17 && 0.243 &0.013 \\
    \#8: -1.20/6.6/22.30 &100 &1 &8.56e+21 &1.74e+13 &1.15e+14 &6.09e+16 &9.74e+13 &3.71e+15 &1.16e+17 && 0.248 &0.016  \\
    \hline   
	\multicolumn{11}{l}{Notes. All the predicted ionic column densities correspond to the ground level except for Fe\,{\tiny II} UV191;} \\
	\multicolumn{11}{l}{The predicted emission-line ratios include contributions from resonance-line doublets C\,{\tiny IV}\,$\lambda\lambda 1548, 1550$ and N\,{\tiny V}\,$\lambda\lambda 1238,1242$.}\\
  \end{tabular}}
\end{table*}

\subsection{DLA Trough and Ly$\alpha$-like Emissions}\label{subsec:Emissions}  

In the \lya\ emission line range, the apparent lack of \lya\ absorption trough is an artifact of comparing them to the prominent broad H$\alpha$ emission detected in the NIR Subaru CISCO data with blueshifted absorption \citep{Aoki06}. 

\begin{figure}
 \centering 
\includegraphics[width=\linewidth]{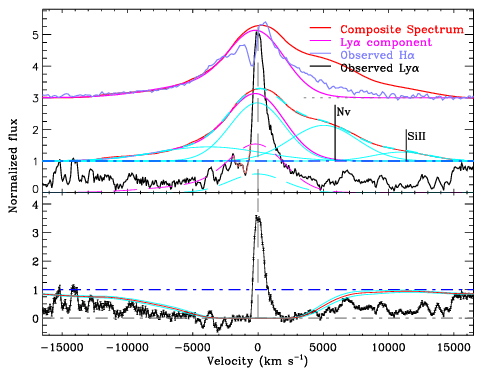}  
\caption{Decomposition of the Ly$\alpha$ spectral region and the DLA trough in velocity space. Top panel shows the similarity between the observed broad emission line profile and the model predicted profile of Ly$\alpha$ in magenta long dashes line which represents the residual flux due to partial covering of the BLR. Bottom panel shows a strong DLA trough with an emission core obtained by subtracting the residual flux from the observed spectrum (black line). The theoretical Voigt profiles for $\log\,\nhi(\cmcl)=21.95\pm 0.05$ predicted by photoionization modeling are shown in red and cyan.} \label{blp}
\end{figure}

Fig.\,\ref{blp} shows the detection of an unusual DLA system where an exceptionally intense \lya\ emission superposed on the trough of the DLA in quasar SDSS J0839+3805. All of the spectra are normalized to unity in the continuum levels (dot-dashed blue horizontal line). 
Here, a single power-law approximation (solid blue line in Fig.\ref{SDSS J0839 SED}) clearly fails on the blue wing of \lya\ emission. Thus, a broken power-law is needed to fit the continuum. We fitted a linear continuum constrained by the EDR flux at two points $\lambda$ = 1157.5 and 1283.6 {\A}, that are relatively uncontaminated with line emission features. The EDR flux is consistent with the maximum measured flux near these two points within uncertainty of 0.05 dex. For simplicity, we neglected the influence of accumulated averaged \lya\ forest absorption on the intrinsic spectrum of the quasar.
The subtraction of broad Ly$\alpha$ emission and the blends with \nv\ and possibly \siii\ are dealt with by means of multiple Gaussian profile decompositions. The decompositions of four Gaussian components (solid cyan line), the broad Ly$\alpha$ (solid magenta line) feature which can be modeled by a combination of two Gaussians of the four components, and observed (black line) spectra in the Ly$\alpha$ spectral region are presented in the upper panel of Fig.\,\ref{blp}. 
The magenta/cyan long dashes line represents the uncovering fractions of the corresponding component.
The Ly$\alpha$ absorptions which are best fitted with a double Gaussian profile in a weak broad base are drawn in coral solid line.
The broad H$\alpha$ emission drawn in light blue solid line that are offset by +2 is overplotted in the figure for comparison. The composite spectrum assumption is justified by the consistency between Ly$\alpha$ subtraction and the observed H$\alpha$ emission.
The profile of the observed Ly$\alpha$ emission, except for the narrow core, is consistent with the magenta long dashes line being scaled by uncovering factors that manifests itself as the residual flux due to partial coverage of the BLR.  

The residual of the broad wings of the Ly$\alpha$ emission is much below the continuum intensity, suggesting strong Ly$\alpha$ absorption. We extracted a strong DLA trough (the lower panel of Fig.\,\ref{blp}) with a prominent Ly$\alpha$ emission located near the center of the trough by subtracting the residual flux due to partial covering effects, from the observed spectrum. 
The over absorptions in the DLA trough, blueward of \lya\,emission may be deduced from extra absorptions such as: BLR clouds, etc. This is beyond the scope of this paper.
Other uncertainties can be introduced by the difference between the intrinsic absorption-free and the composite spectrum, due to data calibration noises, etc.
Considering both the numerous low ionization absorption lines and strong Ly$\alpha$ emission in the observed spectrum, the Voigt profile is consistent, at least qualitatively, with the observed data.

In the extracted DLA trough, there is a strong intermediate width Ly$\alpha$ emission with a rest equivalent width (EW) and FWHM of $14.4\pm1.1${\A}, and $897\pm69\kms$, respectively. The \lya\ line shows an asymmetric profile  having a sharp blueward cutoff and an extended red wing over a range of $\sim$ 2000\kms\ from about $-$500 to 1500\kms\ with respect to systemic velocity.
This velocity range contrary to BAL troughs (from about 500 to $-1500\kms$) shows an opposite trend with the similar velocity width.
If the absorption troughs of certain resonance lines in the spectrum of SDSS J0839+3805 are produced by pure scattering then they must be compensated by an equal amount of emission. We shall tentatively assume that the outflow is axisymmetric (see \citealt{1970ApJ.161L.115S} for a spherically symmetric expanding envelope). Then, the photons scattered from the half of the approching windflow (the near side) will be blueshifted, while scattering from the receding half (the far side) will be redshifted and distributed over a similar frequency interval.
A summation of absorption and scattering form the total line profile that consists of absorption with an emission wing to the red. This profile resemble P Cygni profile (i.e., an undisplaced emission line with a blueshifted absorption component). The Ly$\alpha$ emission and blueshifted absorption lines extend to about the same velocity range ($\sim$ 2000\kms), and together they resemble P Cygni profiles, suggesting self-absorption \citep{2007ApJ.659.250C}.

\begin{figure}
  \centering 
  \includegraphics[width=\linewidth]{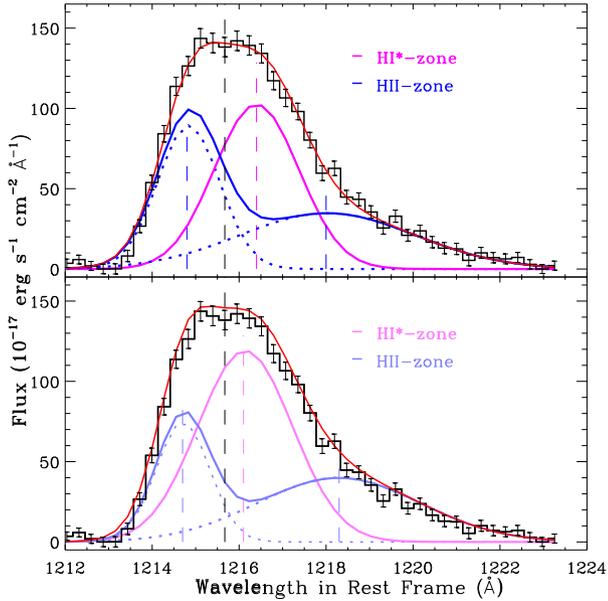}  
  \caption{Three Gaussian profile decompositions of Ly$\alpha$ emission core obtained by subtracting the residual flux from the observed spectrum. Top panel: the intermediate width Ly$\alpha$ emission has been fitted with three Gaussians centered at the rest wavelengths of 1214.8, 1216.4 and 1218.0{\A} with peak values $\sim$ 89.4, 101.9 and 34.7, respectively.  We further assumed that these three Gaussians are divided into two categories based on the emitting region. Bottom panel: another decomposition of Ly$\alpha$ emission with three gaussians centered at the rest wavelengths of 1214.7, 1216.1 and 1218.3{\A} with peak values $\sim$ 74.7, 118.2 and 39.4, respectively.}
\label{3G_dec}
\end{figure}
 
Fig.\,\ref{3G_dec} shows three Gaussian fittings of the Ly$\alpha$ emission. Top panel shows the intermediate width Ly$\alpha$ emission was fitted with three Gaussians of FWHMs $\sim$ 416.8, 564.9, and 1170.9 \kms centered at the rest wavelengths of 1214.8, 1216.4 and 1218.0{\A} with peak values of $\sim$ 89.4, 101.9 and 34.7, respectively. Furthermore, we assumed a priori that these three Gaussians are divided into two categories based on the emitting region; the middle Gaussian profile was produced in \his-zone and a double-peaked profile consisting of the other two Gaussians was produced in \hii-zone (see the next section for an analysis of legitimizing the assumption). 
Note that this fitting is quite subjective and is based upon visual inspection. Bottom panel of Fig.\,\ref{3G_dec} is another decomposition of Ly$\alpha$ emission with three gaussians centered at the rest wavelengths of 1214.7, 1216.1 and 1218.3{\A} with peak values $\sim$ 74.7, 118.2 and 39.4, respectively.

This emitter ought to be seen from other intercepting ions (e.g. \civ, \nv) by resonance line scattering and fluorescent emission due to reprocessing of the nuclear continuum radiation. 
Detailed analysis of the residual intensities in the \civ\ and \nv\ troughs led to the conclusions that there indeed exist significant double-peaked Ly$\alpha$-like emission in \civ\ and \nv\ troughs. In addition to the double-peaked component, the residual intensity in \civ\ trough had revealed a prominent Gaussian component similar to that seen in Ly$\alpha$ emission (i.e., \his-zone component) despite having a relatively reduced line efficiency. 
These results are in contradiction with the profile of Ly$\alpha$ emission from uniform density model as a whole. 

Although the synthetic spectrum from the photoionization models roughly fit the observed spectrum, there are some discrepancies in the model predictions for an absorber with a much smaller radial thickness ($\Delta{R}/R \sim10^{-4}$, here the radial thickness is measured using $\Delta{R} \sim N_{\rm H}/n_{\rm H}$) and extremely low ratio of the \nv\ to Ly$\alpha$ emergent line intensities (see Table\,\ref{tbl-2}). 

\section{A Two-component Model and The Results}\label{sec:Results}

A uniform density model overestimates the ionic column densities of ground and low excitation levels of \feii, and \niii, as well as underestimate \nv\,emission. The predicted emission-line ratios of \nv/Ly$\alpha$ by photoionization simulations from a uniform density model should be less than two percent at most (see Table\,\ref{tbl-2}). We, thus relaxed the assumption of uniform density to get a more physically plausible condition of the absorbing gas. 
The specific steps are: we first fixed the location and density of \his-zone\ to those calculated in preliminary solution for a uniform density model, and then allowed the density of \hii-zone to decrease as the ionization parameter increases. Eventually, this increases the radial thickness of the absorber leading to a negligible influence on the \his-zone behind it. A two-density model of the \hii-zone with $\log\,U=1.0$, $\log\,\dnh(\cmcb)=4.5$, and $\log\,\nh(\cmcl)=23.5$, 
followed by \his-zone behind it with $\log\,\dnh(\cmcb)=6.6$, and $\log\,\nh(\cmcl)=22.25$ with a solar metallicity and turbulence, $v=100 \kms$ could partly   
solve the problem that occurred in a uniform density model, without significantly influencing the following region. 

\subsection{A Turbulent, Near-solar Metallicity, Two-component Model}\label{subsec:Two-component}

To clarify the nature of the two-component model, we generated a grid of refined models in the $U$-$\dnh$-$\nh$ space with the ranges,  $0<\log\,U<1.5$, $4<\log\,\dnh(\cmcb)<9$, and $22.6<\log\,\nh (\cmcl)<24$ in steps of 0.1 dex for all the three parameters with turbulence, $v=100 \kms$ using Cloudy.  

\begin{figure*} 
 \centering
 \includegraphics[width=\linewidth]{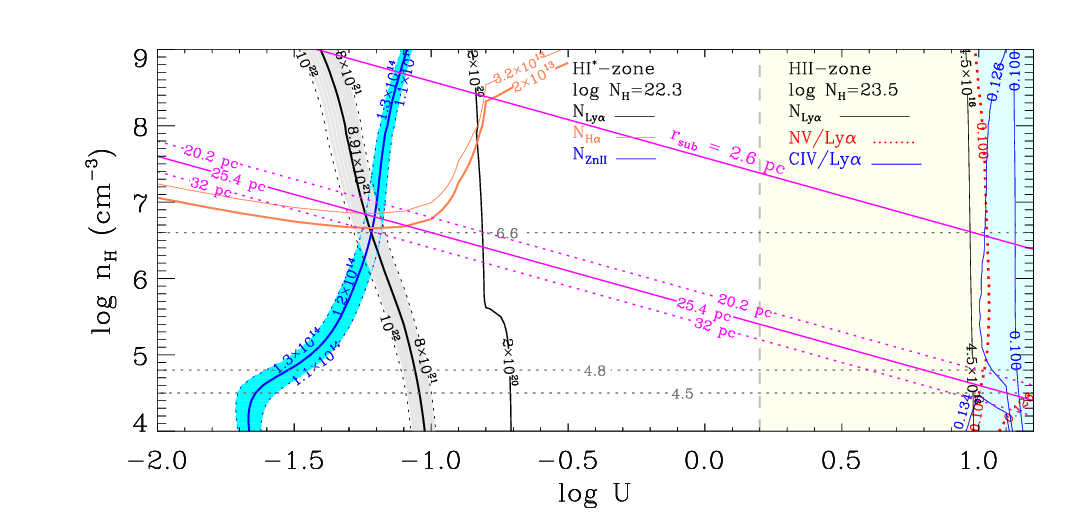} 
 \caption{A two-component model constructed with photoionization simulations by using the combined absorption line and emission-line ratios. The left part of the figure shows a strong DLA absorber constrained by the measured column densities of \znii, and H$\alpha$ in \his-zone. The right part shows the \hii-zone located at a distance very similar to that of the DLA absorber which can at least partly produce the desired \nv\,emission.}
\label{Absorber par} 
\end{figure*}

A two-component model constructed with photoionization modeling by using the combined absorption and emission-line ratios that arise from the same gas is shown in Fig.\,\ref{Absorber par}. 

\begin{figure*} 
 \centering
 \includegraphics[width=0.95\linewidth]{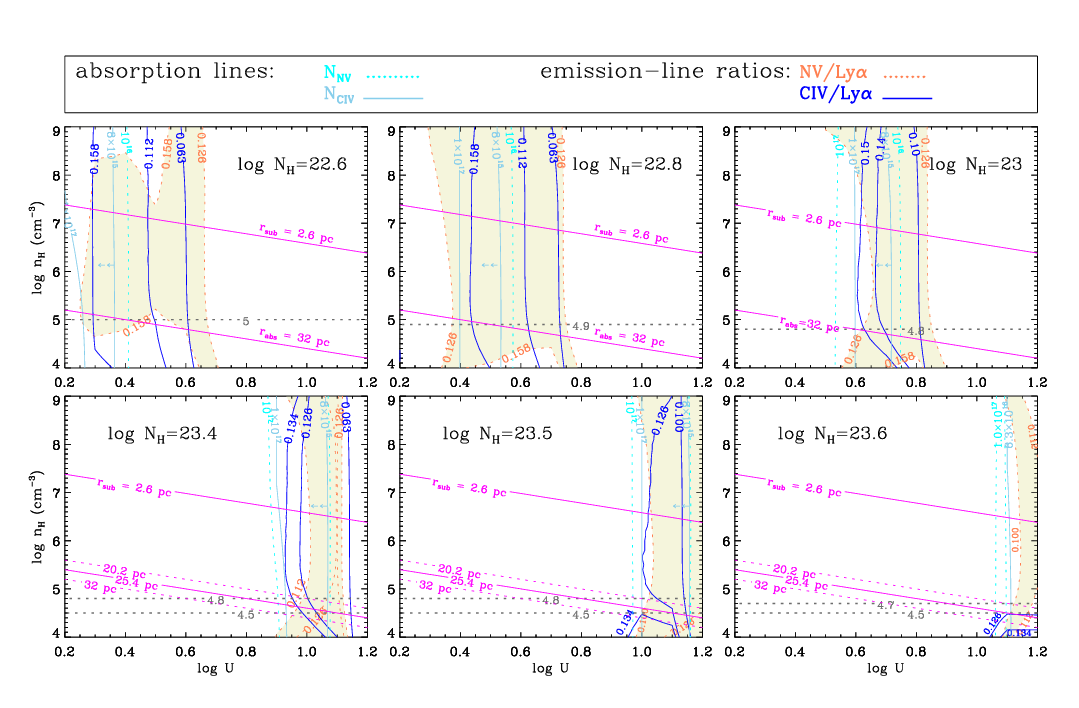} 
 \caption{The model parameters constrained both by the emission-line ratios and absorption lines. The dependence of the predicted emission-line ratios of C\,{\tiny IV}/Ly$\alpha$ and N\,{\tiny V}/Ly$\alpha$ on their ionization parameters and densities are shown. Overplotted on the data are absorption lines of C\,{\tiny IV}, and N\,{\tiny V}. A reasonable range of $\log\,N_{\rm C\,\textsc{\lowercase {IV}}}(\cmcl)$ from $\sim 15.9$ to $17$ were obtained by manual adjustment and iteration by trial and error in order to fit the C\,{\tiny IV} profile.} 
\label{Emission par} 
\end{figure*}

In the figure, the left part shows a strong DLA absorber in \his-zone constrained by the measured column densities of \znii, and H$\alpha$ \citep[also see][for the method used to determine parameters]{2021ApJ.914.13T} and the right part shows a lesser dense gas in \hii-zone located at a distance very similar to that of the DLA absorber which reveals a greatly enhanced emergent emission-line efficiency of \nv. An obvious emergent line intensity of \nv\ is detected in the observed spectrum. The dependency of the emission-line ratios of \civ/Ly$\alpha$ and \nv/Ly$\alpha$ on their ionization parameters and densities predicted by photoionization simulations are shown in Fig.\,\ref{Emission par}. Absorption lines of \civ, and \nv\ are overplotted on the data.  
We found that manual adjustments and iteration by trial and error provide more flexibility in situations where the line profiles are irregular and the blending is potentially severe. This way a reasonable fit to the \civ\ profile can be obtained for a range of $\sim 15.9 - 17$ for $\log\,N_{\rm C\,\textsc{\lowercase {IV}}}(\cmcl)$.

Based on the similar distance reported above, we developed a continuous two-component model for the absorbing cloud. A model consisting of a continuous stratified medium which is divided into two zones; a highly ionized \hii-zone where the clouds are exposed to the full ionizing continuum and a weakly ionized zone called the excited \his-zone behind the \hii-zone which is exposed only to the continuum filtered through the \hii-zone.
Using this model, we performed a series of photoionization equilibrium calculations for a range of model parameters ($\log(U/n_{\rm H}/N_{\rm H})$, turbulent velocity, and metallicity). Here, the turbulent velocity and metallicity were considered as free parameters by considering both the column densities of \feii, and \niii\ from highly excited levels which are sensitive to turbulent velocity, and from ground and relatively low excitation levels which are overestimated by solar metallicity models. The latter gives a clue that the metallicity of the absorbing cloud is probably below that of the solar.  
Table\,\ref{tab:tbl-3} lists the predicted column densities of some typical ions and emission-line ratios of the strong resonance-line doublets \civ\,$\lambda\lambda 1548, 1550$ and \nv\,$\lambda\lambda 1238, 1242$ relative to Ly$\alpha$.
It should be borne in mind, however, that the photoionization analysis and observational constraints on emission lines suffer from large uncertainties.

\begin{landscape}
  \begin{table*}
  \vspace*{20mm}{ 
  \caption{A Two-component Model}\label{tab:tbl-3}
   \hspace*{-50mm}{  
   \resizebox{1.15\linewidth}{!}{
   \begin{tabular}[]{lcccccccccccccccc}
     \hline\hline
     \raisebox{-2ex}{Model} & \multicolumn{7}{c}{H\,{\tiny II}-zone} &&\multicolumn{7}{c}{\raisebox{0ex}{H\,{\tiny I}*-zone}} \\[-1ex]
     \cline{2-8} \cline{10-16} \\[-2.ex]
     & $\log(U/n_{\rm H}/N_{\rm H})$ &Ly$\alpha$ & C\,{\tiny IV} & N\,{\tiny V} &Si\,{\tiny IV} &N\,{\tiny V}/Ly$\alpha^{a}$ &C\,{\tiny IV}/Ly$\alpha^{b}$ && $\log(n_{\rm H}/\nhi)$ & H$\alpha$ & Zn\,{\tiny II} & Fe\,{\tiny II}\,UV191 & Al\,{\tiny III} & N\,{\tiny V}/Ly$\alpha^{a}$ &C\,{\tiny IV}/Ly$\alpha^{b}$ \\[0.ex]
     \hline
     {} & \multicolumn{7}{c}{Turbulent velocity $v$ = 100 km s$^{-1}$, solar metallicity Z=Z$_{\odot}$} &&\multicolumn{7}{c}{Turbulent velocity $v$ = 100 km s$^{-1}$, subsolar metallicity Z=0.5Z$_{\odot}$} \\[.5ex]
     \cline{2-8} \cline{10-16} \\[-2.ex]
     \#1\ldots &\textbf{0.5/5.0/22.8} &2.04e+16 &1.07e+16 &1.80e+16 &2.98e+10 &0.138/0.106 &0.134/0.105  &&\textbf{6.6/22.0} &1.20e+13 &1.09e+14 &7.88e+13 &1.61e+15 &0.004/0.002 &0.082/0.040 \\ 
     \#2\ldots &\textbf{0.7/4.9/23.0} &1.83e+16 &9.35e+15 &1.49e+16 &4.82e+09 &0.134/0.108 &0.132/0.108  &&\textbf{6.6/22.0} &1.38e+13 &1.34e+14 &9.42e+13 &1.38e+15 &0.005/0.003 &0.090/0.044 \\
     \#3\ldots &\textbf{0.9/4.7/23.3} &2.39e+16 &2.52e+16 &2.74e+16 &3.74e+10 &0.117/0.095 &0.133/0.107  &&\textbf{6.6/22.0} &1.21e+13 &1.75e+14 &8.71e+13 &9.82e+14 &0.004/0.002 &0.080/0.039 \\ 
     \#4\ldots &\textbf{0.9/4.8/23.4} &3.72e+16 &1.33e+17 &1.03e+17 &3.26e+14 &0.099/0.078 &0.142/0.103  &&\textbf{6.6/22.0} &1.42e+13 &1.88e+14 &1.03e+14 &8.72e+14 &0.0004/0.0001 &0.070/0.033 \\
     \#5\ldots &\textbf{1.0/4.5/23.4} &2.29e+16 &2.18e+16 &2.44e+16 &1.03e+10 &0.119/0.099 &0.132/0.109  &&\textbf{6.6/22.0} &9.16e+12 &1.83e+14 &6.76e+13 &8.79e+14 &0.004/0.002 &0.075/0.036 \\
     \#6\ldots &\textbf{1.0/4.5/23.5} &3.56e+16 &8.94e+16 &7.42e+16 &1.08e+14  &0.100/0.080 &0.133/0.100  &&\textbf{6.6/22.0} &8.57e+12 &1.92e+14 &6.44e+13 &7.81e+14 &0.0006/0.0002 &0.058/0.027 \\
     \#7\ldots &\textbf{1.0/4.6/23.5} &3.55e+16 &8.93e+16 &7.41e+16 &1.05e+14 &0.098/0.079 &0.132/0.099  &&\textbf{6.6/22.0} &1.07e+13 &1.93e+14 &7.98e+13 &8.06e+14 &0.0008/0.0003 &0.065/0.031 \\
     \#8\ldots &\textbf{1.0/4.7/23.5} &3.55e+16 &8.91e+16 &7.39e+16 &1.03e+14 &0.097/0.079 &0.130/0.098  &&\textbf{6.6/22.0} &1.34e+13 &1.94e+14 &9.86e+13 &8.31e+14 &0.0012/0.0004 &0.071/0.034 \\
     \#9\ldots &\textbf{1.1/4.6/23.6} &3.39e+16 &6.99e+16 &5.95e+16 &2.60e+13 &0.097/0.079 &0.124/0.096  &&\textbf{6.6/22.0} &1.25e+13 &1.99e+14 &9.29e+13 &8.01e+14 &0.0018/0.0007 &0.072/0.034 \\
     \hline
     {} & \multicolumn{7}{c}{Turbulent velocity $v$ = 50 km s$^{-1}$, solar metallicity Z=Z$_{\odot}$} &&\multicolumn{7}{c}{Turbulent velocity $v$ = 50 km s$^{-1}$, subsolar metallicity Z=0.5Z$_{\odot}$} \\[.5ex] 
     \cline{2-8} \cline{10-16} \\[-2.ex]
     \#10\ldots &\textbf{0.9/4.8/23.3}  &2.34e+16 &2.47e+16 &2.70e+16 &3.12e+10 &0.079/0.061 &0.093/0.069 &&\textbf{6.6/22.0} &2.41e+13 &1.83e+14 &7.94e+13 &9.66e+14 &0.003/0.002 &0.093/0.047 \\
     \#11\ldots &\textbf{0.9/4.8/23.4}  &3.67e+16 &1.20e+17 &9.54e+16 &2.53e+14 &0.072/0.052 &0.111/0.073 &&\textbf{6.6/22.0} &2.30e+13 &1.95e+14 &7.64e+13 &8.34e+14 &0.0006/0.0003 &0.076/0.039 \\
     \#12\ldots &\textbf{1.0/4.7/23.5}  &3.50e+16 &8.45e+16 &7.03e+16 &7.25e+13 &0.068/0.051 &0.099/0.068 &&\textbf{6.6/22.0} &2.17e+13 &2.02e+14 &7.29e+13 &7.92e+14 &0.0011/0.0006 &0.077/0.039 \\
     \#13\ldots &\textbf{1.0/4.7/23.6}  &7.54e+16 &1.72e+19 &1.65e+17 &1.07e+16 &0.037/0.027 &0.112/0.069 &&\textbf{6.6/22.0} &1.99e+13 &2.07e+14 &6.77e+13 &6.90e+14 &0.0000/0.0000 &0.048/0.024 \\[1ex] 
     \hline
     {} & \multicolumn{7}{c}{Turbulent velocity $v$ = 100 km s$^{-1}$, solar metallicity Z=Z$_{\odot}$} &&\multicolumn{7}{c}{Turbulent velocity $v$ = 100 km s$^{-1}$, subsolar metallicity Z=0.3Z$_{\odot}$} \\[.5ex] 
     \cline{2-8} \cline{10-16} \\[-2.ex]
     \#14\ldots &\textbf{0.9/4.6/23.3}  &2.39e+16 &2.52e+16 &2.74e+16 &3.75e+10 &0.118/0.096 &0.135/0.108 &&\textbf{6.0/22.0} &1.21e+13 &9.29e+13 &5.55e+13 &7.79e+14 &0.005/0.003 &0.102/0.050 \\
     \#14-1\ldots &\textbf{0.9/4.6/23.3}  &\ldots & & & & & &&\textbf{6.6/22.0} &1.03e+13 &1.05e+14 &5.47e+13 &6.25e+14 &0.003/0.0015 &0.084/0.041 \\
     \#14-2\ldots &\textbf{0.9/4.6/23.3}  &\ldots & & & & & &&\textbf{7.0/22.0}  &1.02e+13 &1.12e+14 &5.42e+13 &4.98e+14 &0.0012/0.0004 &0.055/0.026\\
     \#15\ldots &\textbf{1.0/4.5/23.5}  &3.56e+16 &8.94e+16 &7.42e+16 &1.08e+14 &0.0998/0.080 &0.133/0.100 &&\textbf{6.0/22.0} &1.09e+13 &1.11e+14 &5.07e+13 &6.28e+14 &0.002/0.001 &0.090/0.044 \\
     \#15-1\ldots &\textbf{1.0/4.5/23.5}&\ldots & & & & & &&\textbf{6.6/22.0}  & 9.20e+12 &1.17e+14 &5.01e+13 &5.05e+14 &0.0004/0.0001 &0.068/0.033\\
     \#15-2\ldots &\textbf{1.0/4.5/23.5}&\ldots & & & & & &&\textbf{7.0/22.0}  &9.06e+12 &1.21e+14 &4.98e+13 &4.19e+14 &0.0001/0.0000 &0.042/0.019\\
     \#16\ldots &\textbf{1.1/4.5/23.6}  &3.40e+16 &6.99e+16 &5.95e+16 &2.65e+13 &0.098/0.080 &0.126/0.097 &&\textbf{6.0/22.0} &1.32e+13 &1.12e+14 &6.12e+13 &6.59e+14 &0.005/0.003 &0.104/0.051 \\
     \#16-1\ldots &\textbf{1.1/4.5/23.6} &\ldots & & & & & &&\textbf{6.6/22.0} &1.08e+13 &1.21e+14 &5.84e+13 &5.01e+14 &0.0010/0.0004 &0.077/0.037\\
     \#16-2\ldots &\textbf{1.1/4.5/23.6} &\ldots & & & & & &&\textbf{7.0/22.0} &1.06e+13 &1.23e+14 &5.79e+13 &4.28e+14 &0.0003/0.0001 &0.049/0.023\\
     \hline
     {} & \multicolumn{7}{c}{Turbulent velocity $v$ = 100 km s$^{-1}$, subsolar metallicity Z=0.8Z$_{\odot}$} &&\multicolumn{7}{c}{Turbulent velocity $v$ = 100 km s$^{-1}$, subsolar metallicity Z=0.5Z$_{\odot}$} \\[.5ex] 
     \cline{2-8} \cline{10-16} \\[-2.ex]
     \#17\ldots &\textbf{0.7/5.0/23.2} &3.23e+16 &5.05e+16 &4.32e+16 &7.67e+11 &0.108/0.082 &0.149/0.105 &&\textbf{6.6/21.9} &1.28e+13 &1.20e+14 &9.92e+13 &1.10e+15 &0.0026/0.001 &0.079/0.038 \\
     \#18\ldots &\textbf{0.8/4.9/23.3} &3.10e+16 &4.25e+16 &3.77e+16 &2.59e+11  &0.104/0.081 &0.135/0.10 &&\textbf{6.6/21.9} &1.22e+13 &1.30e+14 &9.62e+13 &1.01e+15 &0.003/0.0014 &0.079/0.038 \\
     \#19\ldots &\textbf{0.9/4.8/23.4} &3.00e+16 &3.70e+16 &3.38e+16 &8.35e+10 &0.101/0.080 &0.126/0.096 &&\textbf{6.6/22.0} &1.45e+13 &1.75e+14 &1.04e+14 &9.96e+14 &0.0032/0.0015 &0.080/0.039 \\
     \#19-1\ldots &\textbf{0.9/4.8/23.4}  &\ldots & & & & & &&\textbf{6.6/21.9} &1.16e+13 &1.39e+14 &9.26e+13 &9.36e+14 &0.0032/0.0016 &0.080/0.039 \\
     \#20\ldots &\textbf{1.0/4.8/23.5} &2.92e+16 &3.26e+16 &3.06e+16 &2.72e+10 &0.098/0.079 &0.117/0.093 &&\textbf{6.6/22.0} &1.71e+13 &1.85e+14 &1.22e+14 &9.58e+14 &0.004/0.002 &0.087/0.042 \\
     \#20-1\ldots &\textbf{1.0/4.8/23.5} &\ldots & & & & & &&\textbf{6.6/21.9} &1.37e+13 &1.46e+14 &1.08e+14 &9.07e+14 &0.004/0.002 &0.087/0.042 \\
     \#21\ldots &\textbf{1.1/4.7/23.6} &2.85e+16 &2.87e+16 &2.78e+16 &9.42e+09 &0.096/0.080 &0.112/0.091 &&\textbf{6.6/22.0} &1.59e+13 &1.92e+14 &1.15e+14 &8.97e+14 &0.004/0.002 &0.089/0.043 \\
     \hline
     {} & \multicolumn{7}{c}{Turbulent velocity $v$ = 100 km s$^{-1}$, subsolar metallicity Z=0.5Z$_{\odot}$} &&\multicolumn{7}{c}{Turbulent velocity $v$ = 100 km s$^{-1}$, subsolar metallicity Z=0.5Z$_{\odot}$} \\[.5ex] 
     \cline{2-8} \cline{10-16} \\[-2.ex]
     \#22\ldots &\textbf{0.9/4.7/23.5} &3.36e+16 &2.61e+16 &2.24e+16 &2.73e+09 &0.088/0.070 &0.110/0.084 &&\textbf{6.6/22.0} &1.12e+13 &1.61e+14 &8.05e+13 &1.06e+15 &0.002/0.001 &0.075/0.036 \\
     \#23\ldots &\textbf{0.9/4.8/23.5} &3.36e+16 &2.61e+16 &2.24e+16 &2.73e+09 &0.087/0.069 &0.109/0.083 &&\textbf{6.6/22.0} &1.40e+13 &1.60e+14 &9.95e+13 &1.11e+15 &0.003/0.001 &0.081/0.040 \\
     \#24\ldots &\textbf{1.0/4.7/23.6} &3.34e+16 &2.27e+16 &2.03e+16 &9.98e+08 &0.084/0.068 &0.101/0.079 &&\textbf{6.6/22.0} &1.30e+13 &1.70e+14 &9.36e+13 &1.01e+15 &0.003/0.001 &0.082/0.040\\
     \#25\ldots &\textbf{1.0/4.8/23.6} &3.34e+16 &2.27e+16 &2.03e+16 &9.99e+08 &0.083/0.067 &0.099/0.078 &&\textbf{6.6/22.0} &1.62e+13 &1.70e+14 &1.15e+14 &1.05e+15 &0.003/0.002 &0.088/0.043\\
     \#26\ldots &\textbf{1.1/4.6/23.7} &3.34e+16 &2.07e+16 &1.90e+16 &4.27e+08 &0.082/0.067 &0.095/0.076 &&\textbf{6.6/22.0} &1.19e+13 &1.79e+14 &8.67e+13 &9.22e+14 &0.0028/0.0014 &0.082/0.040 \\
     \#27\ldots &\textbf{1.1/4.7/23.7} &3.34e+16 &2.07e+16 &1.90e+16 &4.26e+08 &0.080/0.066 &0.094/0.075 &&\textbf{6.6/22.0} &1.49e+13 &1.80e+14 &1.07e+14 &9.58e+14 &0.0032/0.0016 &0.089/0.043 \\
     \hline\hline
   \multicolumn{16}{l}{Notes. $^\textbf{a}$ the predicted emission-line ratios of N\,{\tiny V}\,$\lambda\lambda 1238, 1242$ doublets relative to Ly$\alpha$; $^\textbf{b}$ the predicted emission-line ratios of C\,{\tiny IV}\,$\lambda\lambda 1548, 1550$ doublets relative to Ly$\alpha$.}\\
   \end{tabular} 
   }}}
  \end{table*}
\end{landscape} 

Among the models, model \#19-1 consisting of a \hii-zone with $\log\,U=0.9$, $\log\,\dnh(\cmcb)=4.8$, $\log\,\nh(\cmcl)=23.4$, and Z = 0.8Z$_{\odot}$, followed by \his-zone behind it with $\log\,\dnh(\cmcb)=6.6$, $\log\ \nhi =21.9$, and Z = 0.5Z$_{\odot}$ with the same turbulent velocity of $v=100 \kms$ seemed in best agreement with the observed spectrum, and thus it was accepted as our final solution.

\begin{figure*} 
 \centering
 \includegraphics[width=1.0\linewidth]{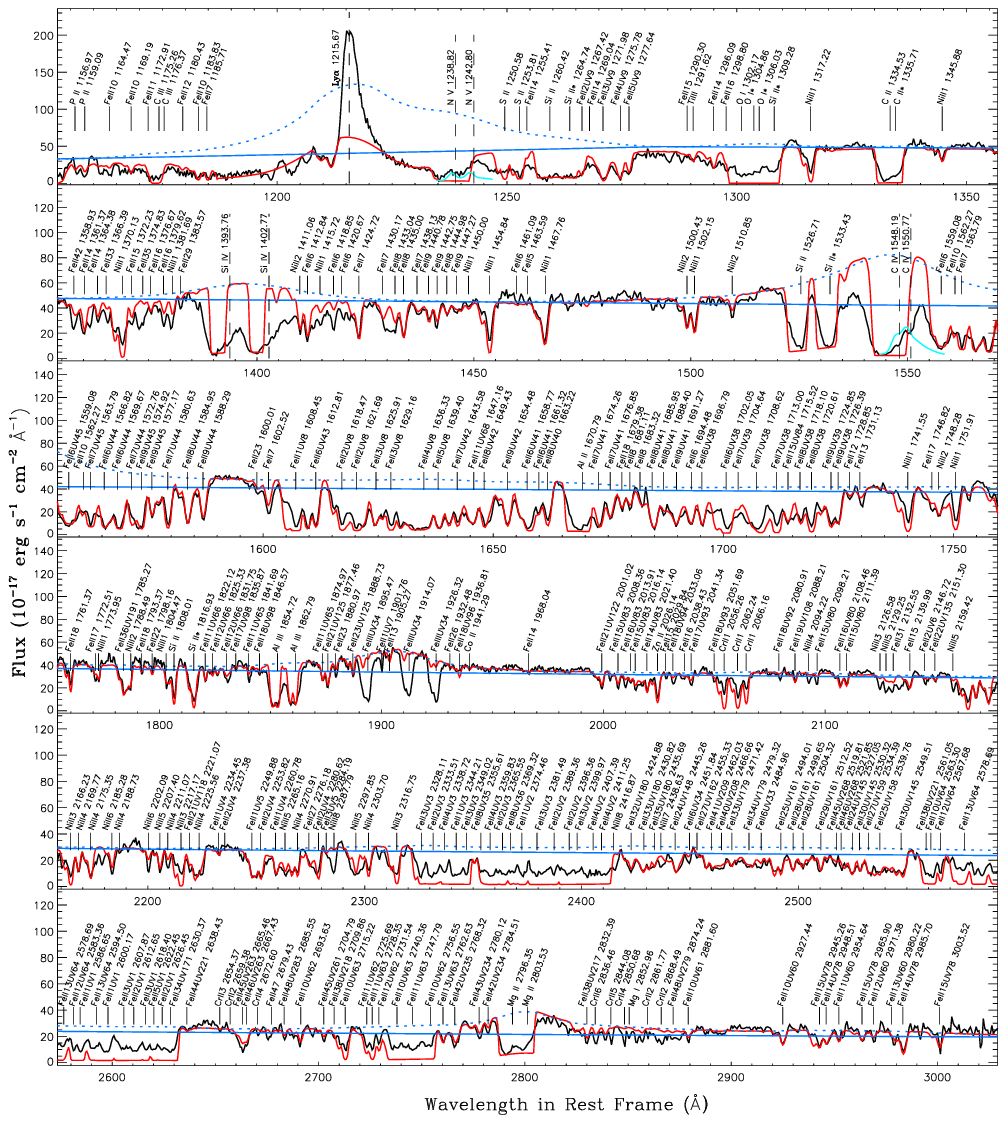}
 \caption{The synthetic spectrum of photoionization modeling. The red line represents the synthetic absorption spectrum constructed from Cloudy model based on a continuous stratified medium. 
The synthetic emission spectra of the strong resonnace doublets, C\,{\tiny IV}\,$\lambda\lambda 1548, 1550$ and N\,{\tiny V}\,$\lambda\lambda 1238, 1242$ predicted by Cloudy model are drawn in solid cyan lines superimposed on the residual intensities due to partial covering effects at the bottom of the absorption lines. The gray long-dashed vertical lines correspond to the rest-frame wavelength of each emission line.}
\label{Absorber modelp} 
\end{figure*}

The optimized synthetic spectra constructed from Cloudy models using the fitting method used in our previous work \citep{2016ApJ.819.99S} are plotted in Fig.\,\ref{Absorber modelp}.
The red line represents the synthetic absorption spectrum constructed from Cloudy model.
The dotted and solid blue lines show the reddened EDR composite spectrum and power-law continuum, respectively.
The synthetic emission spectra of strong resonance line doublets \civ\,$\lambda\lambda 1548, 1550$, and \nv\,$\lambda\lambda 1238, 1242$ 
predicted by Cloudy model are drawn in cyan solid lines superimposed on the residual intensities due to partial covering effects at the bottom of the absorption lines. 
Part of the identified absorption lines (not all of them for clarity) are labeled in the figure. All the line wavelengths are vacuum values. We created the notation for absorption lines of \feii, \niii\ and \crii\ just by adding the number of low-lying energy levels following by the ion stage of the element symbol, and considering that many lines of transitions from highly excited levels of \feii, \niii, \crii\ are heavily blended. All other symbols have standard meanings.
The synthetic spectra successfully fit, at least qualitatively, the observations.

We would like to emphasize that certain atomic data for \feii\ from \citet{Verner99} were substituted with three other datasets of \feii\ to enhance the synthetic spectra (refer to Appendix~\ref{comparison} for a detailed description). 

\subsection{Geometrical Configuration of the Absorber}\label{subsec:Location} 
Here, we estimated the position and radial range of the UV absorber in SDSS J0839+3805. The location of the intrinsic absorber (defined by distance $R_{\rm{abs}}$) with respect to the central source can be constrained by the ionization parameter, ($U= Q_{\rm H}/4 \pi R_{\rm{abs}}^2 \dnh c$), which gives the ratio of number densities between hydrogen ionizing photons ($Q_{\rm H}$) at the face of the absorber and hydrogen in all forms (\dnh). Here, $c$ is the speed of light. The value of $Q_{\rm H}$ can be evaluated using the SED of the ionizing continuum. Based on Cloudy models, the distance $R_{\mathrm{abs}}$ as a function of $U$, $\dnh$ and $Q_{\rm H}$ can be easily derived. Our uniform density model yields a radial distance of $R_{\mathrm{abs}} \approx$ 32 pc from the nucleus to the absorbing clouds.
Furthermore, the revised two-component model can refine constraints on the radial span of the absorber and on $R_{\rm{abs}}$. We can place a lower limit on the radial thickness of a contiguous absorber in our Cloudy model using $\Delta{R} \sim N_{\rm H}/n_{\rm H}$.

The final solution of model \#19-1 yields a radial span of the absorber $\sim$ 1.3 pc, originating at a radial distance of $R_{\mathrm{abs}}$ = 22.7 pc. From the model group with metallicities, Z = 0.8/0.5 Z$_{\odot}$, listed in Table\,\,\ref{tab:tbl-3}, we conservatively estimated $R_{\rm{abs}}$ and the radial span of the absorber as $22.7^{+2.7}_{-2.5}$ pc, and 1.3$\sim$2.6 pc, respectively.

To estimate the relative position of the absorber among nuclear environments, we constrained the other characteristic sizes of the object. 
Using the \rl\ relationship \citep[see equation (2) of][]{2009ApJ.697.160B}, we can estimate the size of BLR. 
We measured an extinction-corrected luminosity at 5100\A\ for an extinction, $E(B - V)=0.03$, which gives $\lambda L_{\lambda}(5100)=1.90\times 10^{46}\ergs$ that corresponds to a BLR of size, $R_{\rm{BLR}}$ = 0.44 pc.

The dust sublimation radius which can act as the inner edge of the torus is given by $R_{\rm{sub}}=1.3L^{1/2}_{uv,46}T^{-2.8}_{1500}$ pc \citep[see equation (5) of][]{1987ApJ.320.537B}. Here, $L_{uv,46}$ is the total ultraviolet luminosity in units of $10^{46} \ergs$ and $T_{1500}$ is the grain evaporation temperature in units of 1500 K. For SDSS J0839+3805, assuming a dust temperature of 1500 K, we obtained $R_{\rm{sub}}\approx$ 2.6 pc using the model result $L_{uv,46}$ = 4.02. These values can be used to estimate the torus size which is usually 5-10 times larger than $R_{\rm{sub}}$ \citep{2006ApJ.648L.101E}.

We roughly measured the FWHMs of the \lya\ emission in the DLA trough and the broad H$\alpha$ emission as $\sim 897$ and $6700\kms$, respectively. Assuming that the Ly$\alpha$ emitter is similar to the BLR, $R_{\rm{BLR}}\propto\rm{FWHM_{BLR}}^{2}$ and using the FWHMs of the Ly$\alpha$ and H$\alpha$ (which represent BLR emission), we found that the Ly$\alpha$ emitter is about (6700/897)$^2$ times farther than the BLR which locates at a distance of $R_{\rm{Ly\alpha}}\approx$ 24.5 pc from the nucleus.
This is consistent with the result $\sim$21.5 pc estimated by means of the measurements of \lya\ and H$\alpha$ made by \citet{Aoki06} within uncertainty of 0.1 dex between the two measure. The FWHMs of \lya\ and H$\alpha$ are $930\pm25$ and $6500\pm50$, respectively \citep[see Table\, 2 of][]{Aoki06}.
The similar distance of the Ly$\alpha$ emitter and the UV absorber, $R_{\mathrm{abs}}$ from the nucleus suggests that the DLA absorber is also responsible for the Ly$\alpha$ emission.
This result is supported by the characteristics of Ly$\alpha$-like emissions (e.g., \civ, \nv) predicted by our simulations, which align reasonably well with observations. 

The observed Ly$\alpha$ emission exhibits a rest-frame EW of approximately 14.4{\A}. In contrast, the predicted Ly$\alpha$ EW, assuming complete absorption of the ionizing continuum radiation by optically thick BLR clouds \citep{1997bk.Peterson}, is approximately 580{\A}.
This yield a global covering factor of the Ly$\alpha$ emitter to be around 2.5\%, which means that an optically thick cloud would absorb about 2.5\% of the ionizing continuum photons and create an equal amount of Ly$\alpha$ photons via recombination \citep{1979ApJ.233L.91K,2006agna.book.O} towards our line of sight. This covering factor corresponds to a toroidal structure of scale height $\sim$ 1.2 pc that is located at a distance of $\sim$ 23 pc from the nucleus. 
Half of the scale height can therefore be used to set an upper limit on the vertical thickness of the absorbing gas. This is consistent with values estimated based on partial coverage analysis. The partial coverage of the BLR indicates that the absorber has a size that is comparable to, or less than the largest projected size of the BLR.

\section{Summary and Discussion}\label{sec:Discussion}

In our study, we have discovered a strong DLA absorber that is roughly cospatial with the dusty torus in the quasar SDSS J0839+3805 located at redshift, $z = 2.315$. Additionally, we have identified an exceptionally intense \lya\ emission with a rest EW $\sim 14.4\pm1.1${\A} and a $\mathrm{FWHM} \sim 897\pm69\kms$, intermediate between the characteristics of classic broad and narrow emission lines.
This absorber is nearly dust-free and acts as a natural coronagraph and manifests itself in both absorption- and emission-lines. 
Extensive photoionization modeling reveals a turbulent, near-solar metallicity absorber, where the main characteristic of the quasar spectrum can be successfuly reproduced by a two component model. 

Our model consists of a diffuse lower density \hii-zone with $\log\,U=0.9$, $\log\,\dnh(\cmcb)=4.8$, $\log\,\nh(\cmcl)=23.4$, and Z = 0.8Z$_{\odot}$, followed by a denser \his-zone behind it with $\log\,\dnh(\cmcb)=6.6$, $\log\ \nhi=21.9$, and Z = 0.5Z$_{\odot}$ with turbulent velocity, $v=100 \kms$. This absorber has a toroidal structure which is roughly collocated with the dusty torus and it has an upper limit of $\sim0.6$ pc on its vertical thickness.
The conclusive results are grounded in the following pieces of evidence:

\begin{enumerate}
\item DLAs with partial coverage: The residual intensity in the DLA trough, influenced by partial covering from the BLR, is determined by the covering factor measured using the residual intensity at the lowest points of the representative absorption lines of \cii\,$\lambda1334$, \siii\,$\lambda1526$, \civ\,$\lambda1548$, and \mgii\,$\lambda2796$ which are expected to be completely saturated. In any case, partial coverage can constrain the relative sizes of the absorbers and nuclear emission sources, placing a lower limit on the transverse size of the absorber.
\item Non-negligible global covering factor: The observed rest EW of $\sim$ 14.4{\A} of the Ly$\alpha$ emission yields a global covering factor of the Ly$\alpha$ emitter to be around 2.5\%. This covering factor corresponds to a toroidal structure of scale height $\sim$ 1.2 pc that is located at a distance of $\sim$ 23 pc from the nucleus. Half of the scale height can be regarded as the vertical thickness of the absorbing gas which is consistent with estimates from partial coverage analysis.
\item Axisymmetric and self-shielding clouds: The extracted \lya\ emission, which has skewed and redshifted wing (from $\sim -500$ to 1500\kms), contrary to blueshifted BAL troughs (from $\sim$ 500 to $-1500\kms$) which extends to the similar velocity width, suggests an axisymmetric and self-shielding clouds. This is reinforced by the lack of spectral variability, which indicates that the cloud is either far from the source or is being continually replenished, and is azimuthally symmetric \citep{2008ApJ.675.985G}.
\item Collocated with the dusty torus: The location of the Ly$\alpha$ emitter ($R_{\rm{Ly\alpha}}\approx$ 21.5$-$24.5 pc), derived from the widths of the \lya\,and H$\alpha$\,emissions, is consistent with that of the DLA absorber (located at $R_{\rm{abs}} = 22.7$ pc with a radial span of 1.3$\sim$2.6 pc), indicating a shared origin of them from a region that is collocated with the dusty torus.
\item Prediction of emission-line ratios: The photoionization model of the DLA absorber gives emission-line ratios that are close to the observed values of the broad emission lines (\civ/\lya$\sim$ 0.08$-$0.14, \nv/\lya $\sim$ 0.2).
\item The emission-line ratios suggest an intrinsic origin: The line ratios of \civ/\lya\ and \nv/\lya, either measured from the observed spectrum or predicted by our models, are consistent with the mean ratios of \civ/\lya = 0.21 $\pm$ 0.09 that were measured for four local Seyfert 2 galaxies by \citet{1986ApJ.300.658F} and \nv/\lya\ ranging from 0.03 to 0.3 measured in most Seyfert 2 galaxies \citep[see][]{1998ApJ.498L.93D}. In contrast, the non-AGN spectrum has intensity ratios of \civ/\lya\ less or equal to 0.02 \citep[see][]{2003ApJ.588.65S}.
This is also consistent with the finding of \citet{1992ApJ.398.476F} that the BAL QSOs show excess \nv\ emission relative to the QSO population as a whole. 
\end{enumerate}

Taking all these points into account, we conclude that the DLA absorber that is causing the absorptions is also responsible for the Ly$\alpha$-like emissions, and is collocated with the dusty torus. 
If the gas lies along our line of sight to the nucleus, the absorption is observed; if the gas lies out of our line of sight, but still within instrumental angular resolution, only the emission which is attributed to the reprocessing of the central continuum radiation can be detected.
The most plausible scenario is that the dusty torus is inherently geometrically thick, likely a consequence of gas inflow from galactic scales to the black hole, propelled by gravitational torques during its gas-rich phase \citep{2012MNRAS.420.320H}. Thus, this torus can serves as a fuel reservoir, establishing a connection between the outer accretion disk and its host galaxy, thereby regulating the fueling from the torus through a dust and gas flow cycle.
The evaporative absorber (outflow) is being continually replenished in a toroidal structure that can support a geometrically thick  long-lived torus with internal turbulence by interacting with inflow \citep{2011ApJ.741.29D,2012ApJ.758.66W}.

Although the synthetic spectra constructed using the code CLOUDY successfully fit the observations as a whole, there exist some obvious discrepancies such as: significant residuals in absorption troughs of \feii\,multiplets UV 1, UV 2, UV 3, UV 62 and 63 near 2600{\A}, 2382{\A}, 2344{\A}, and 2750{\A}; the observed absorptions of \feiii\,UV 34 triplet are much stronger than predictions from Cloudy photoionization simulations, etc.
The predicted \feii\ emissions from the absorber in the photoionization calculations are much less than the residual fluxes of \feii\ (not shown in Fig.\,\ref{Absorber modelp}). If the residual fluxes are due to \feii\ emission from the BLR leaking through the absorber with covering factor similar to those of \mgii\ ($\sim$0.74), then such high covering factor would result in too strong broad \feii\ emission to credible.
The photoionization models of the BLR in quasars still struggle to account for the observed intensities of \feii.
Our study implies that the strong \feii\ emission emitted from outside of the BLR. This result is consistent with the finding of \citet{2009AJ.137.3548P} that the location of the \feii\ emission region is located outside the BLR (the so-called intermediate line region or ILR), which was confirmed by \feii\ reverberation \citep{2013ApJ.769.128B}. 
\feiii\,UV 34 absorptions are another problem. The unusual prominent three absorption lines $\lambda\lambda 1895.5,1914.1,1926.3$ are \feiii\,absorption by the triplet UV 34 from a low-lying metastable level at 3.73 eV. A possible population mechanism for \feiii\,UV 34 is that the upper level of \feiii\,UV 34 is populated by Ly$\alpha$ pumping \citep{Johansson00}, and subsequent cascade to low-lying metastable levels. But in reality the population mechanism for \feiii\,UV 34 absorptions cannot be considered solved, and may not be well handled in code Cloudy, leading to a much lower \feiii\,UV 34 absorptions predicted by Cloudy simulations.
Other explanations for these discrepancies are: the lines missing from the atomic line list or the available atomic data are likely inaccurate, and the model could be still simplified for the actual physical conditions of the absorber, especially for high-ionization lines, etc.
 
This object is viewed nearly edge-on and exhibits partial coverage to the BLR, thus there is a high possibility that it is viewed through the disk wind, across the BLR clouds, as well as infalling gas near and likely originated inside the dust sublimation radius. The influence of all the flows on the line absorption profile are neglected in this paper.

Finally, we provide a concise discussion on the limitations of absorption line analysis based on low-resolution spectra. Our confidence in the absorption line measurements is supported by the following reasons:

\begin{enumerate}
\item The predicted population of H$\alpha$ is consistent with  the value derived by \citet{Aoki06} where they used the curve of growth which was developed to deal with low-resolution data \citep{1986ApJ.304.739J}.
\item For quasars having high-quality and low-resolution data, as for example SDSS J170542.91+354340.2, absorption line measurements from low-resolution SDSS (R $\sim$ 2000) data and high-resolution VLT/UVES (R $\sim$ 50000) data yield very close column densities, with deviations within the predicted measurement uncertainties of ~0.1 dex \citep[see Table\, 4 of][]{2017ApJ.835.218P}.
\item For SDSS J0839+3805, over 200 absorption lines can be verified, including some very common multiplets like \feii. The fitted profiles for low-ionization lines such as \niii\,$\lambda 1455$ give FWHM of (225 $\pm 13)\kms$, and (252 $\pm 60)\kms$ with deconvolved FWHMs of 168 \kms and 204 \kms (applying the average resolution R $\sim$ 2000), respectively. The line profiles are marginally resolved and have similar wide low-ionization absorptions to those of SDSS J170542.91+354340.2. 
\end{enumerate}

Thus, absorption line measurements can be conducted reliably, with conservative uncertainties of $\sim$0.2 dex, even when the data resolution is low.

\section*{Acknowledgments}
 
We thank K. Aoki for sending the near-infrared spectra data of SDSS J0839+3805 obtained with CISCO on the Subaru telescope. We are grateful to the referee for careful reading as well as constructive comments and suggestions that helped to clarify the paper. This work was supported by the National Key R\&D Program of China (Grant No. 2022YFC2807300), the Space Debris Research Project, China (Grant No. KJSP2020010102), and the Shanghai Natural Science Foundation (Grant No. 22ZR1481200, 21ZR1469800). This work was partially supported by a program of the Polish Ministry of Science under the title 'Regional Excellence Initiative' (project no. RID/SP/0050/2024/1). SHZ acknowledges support from the National Natural Science Foundation of China (Grant No. 12173026), the Shanghai Science and Technology Fund (Grant No. 23010503900), the Program for Professor of Special Appointment (Eastern Scholar) at Shanghai Institutions of Higher Learning and the Shuguang Program (23SG39) of the Shanghai Education Development Foundation and Shanghai Municipal Education Commission. 

\section*{Data Availability}

The spectroscopic data underlying this article are publicly available in SDSS data bases. The atomic data of \feii\ for four datasets are publicly available in the release of CLOUDY. 
The derived data generated in this research will be shared on reasonable request to the first author.




\begin{thebibliography}{}

\bibitem[Aoki \etal(2006)]{Aoki06} Aoki K., Iwata I., Ohta K., Ando M., Akiyama M., Tamura N.,\ 2006, ApJ, 651, 84$-$92
\bibitem[Arav \etal(2001)]{Arav01} Arav N., de Kool M., Korista K.~T. \etal,\ 2001, ApJ, 561, 118
\bibitem[\protect\citeauthoryear{Arav et al.}{2018}]{2018ApJ.857.60A} Arav N., Liu G., Xu X., Stidham J., Benn C., Chamberlain C., 2018, ApJ, 857, 60
\bibitem[Barlow \& Sargent(1997)]{Barlow97} Barlow T.~A., Sargent W.~L.~W., 1997, AJ., 113, 136
\bibitem[Barnes \& Haehnelt(2009)]{2009MNRAS.397.511B} Barnes L.~A., Haehnelt M.~G.,\ 2009, \mnras, 397, 511
\bibitem[\protect\citeauthoryear{Barth et al.}{2013}]{2013ApJ.769.128B} Barth A.~J., Pancoast A., Bennert V.~N., Brewer B.~J., Canalizo G., Filippenko A.~V., Gates E.~L. et al., 2013, ApJ, 769, 128
\bibitem[Barvainis(1987)]{1987ApJ.320.537B} Barvainis R.,\ 1987, \apj, 320, 537
\bibitem[Bentz et al.(2009)]{2009ApJ.697.160B} Bentz M.~C., Peterson B.~M., Netzer H., Pogge R.~W., Vestergaard M.,\ 2009, \apj, 697, 160
\bibitem[Cashman et al.(2017)]{2017ApJS.230.8C} Cashman F.~H., Kulkarni V.~P., Kisielius R., Ferland G.~J., Bogdanovich P.,\ 2017, \apjs, 230, 8
\bibitem[\protect\citeauthoryear{Choi et al.}{2022}]{2022ApJ.937.74C} Choi H., Leighly K.~M., Terndrup D.~M., Dabbieri C., Gallagher S.~C., Richards G.~T., 2022, ApJ, 937, 74
\bibitem[Clavel et al.(1991)]{1991ApJ.366.64C} Clavel J., Reichert G.~A., Alloin D. et al.,\ 1991, \apj, 366, 64
\bibitem[Collin-Souffrin \etal(1986)]{Collin-S86} Collin-Souffrin S., Dumont S., Joly M., Pequignot D., 1986, A\&A, 166, 27
\bibitem[\protect\citeauthoryear{Crenshaw \& Kraemer}{2007}]{2007ApJ.659.250C} Crenshaw D.~M., Kraemer S.~B., 2007, ApJ, 659, 250
\bibitem[de Kool \etal(2002)]{deKool02} de Kool M., Becker R.~H., Arav N., Gregg M.~D., White R.~L., 2002, ApJ, 570, 514
\bibitem[\protect\citeauthoryear{de Kool \& Begelman}{1995}]{1995ApJ.455.448D} de Kool M., Begelman M.~C., 1995, ApJ, 455, 448
\bibitem[Dey et al.(1998)]{1998ApJ.498L.93D} Dey A., Spinrad H., Stern D., Graham J.~R., Chaffee F.~H.,\ 1998, \apjl, 498, L93
\bibitem[Dorodnitsyn et al.(2011)]{2011ApJ.741.29D} Dorodnitsyn A., Bisnovatyi-Kogan G.~S., Kallman T.,\ 2011, \apj, 741, 29
\bibitem[Elitzur \& Shlosman(2006)]{2006ApJ.648L.101E} Elitzur M., Shlosman I.,\ 2006, \apj, 648, L101
\bibitem[Fathivavsari(2020)]{2020ApJ.888.85F} Fathivavsari H.,\ 2020, \apj, 888, 85
\bibitem[\protect\citeauthoryear{Fathivavsari et al.}{2017}]{2017MNRAS.466L.58F} Fathivavsari H., Petitjean P., Zou S., Noterdaeme P., Ledoux C., Kr{\"u}hler T., Srianand R., 2017, MNRAS, 466, L58
\bibitem[Fedchak \& Lawler(1999)]{Fedchak99} Fedchak J.~A., Lawler J.~E., 1999, ApJ, 523, 734
\bibitem[Ferland et al.(2017)]{2017RMxAA.53.385F} Ferland G.~J., Chatzikos M., Guzm{\'a}n F. et al.,\ 2017, \rmxaa, 53, 385
\bibitem[Ferland \& Osterbrock(1986)]{1986ApJ.300.658F} Ferland G.~J., Osterbrock D.~E.,\ 1986, \apj, 300, 658
\bibitem[Finley et al.(2013)]{2013A&A.558A.111F} Finley H., Petitjean P., P{\^a}ris I. et al.,\ 2013, \aap, 558, A111
\bibitem[Fitzpatrick \& Massa(2007)]{2007ApJ.663.320F} Fitzpatrick E.~L., Massa D.,\ 2007, \apj, 663, 320
\bibitem[Francis et al.(1992)]{1992ApJ.398.476F} Francis P.~J., Hewett P.~C., Foltz C.~B., Chaffee F.~H.,\ 1992, \apj, 398, 476
\bibitem[Gabel \etal(2005a)]{Gabel05} Gabel J.~R., Arav N., Kaastra J.~S. \etal,\ 2005a, ApJ, 623, 85
\bibitem[\protect\citeauthoryear{Gabel et al.}{2005b}]{2005ApJ.631.741G} Gabel J.~R., Kraemer S.~B., Crenshaw D.~M., George I.~M., Brandt W.~N., Hamann F.~W., Kaiser M.~E. et al., 2005b, ApJ, 631, 741
\bibitem[Gibson et al.(2008)]{2008ApJ.675.985G} Gibson R.~R., Brandt W.~N., Schneider D.~P., Gallagher S.~C.,\ 2008, \apj, 675, 985
\bibitem[Gordon \etal(2003)]{Gordon03} Gordon K.~D., Clayton G.~C., Misselt K.~A., Landolt A.~U., Wolff M., 2003, ApJ, 594, 279
\bibitem[\protect\citeauthoryear{Hall et al.}{2002}]{2002ApJS.141.267H} Hall P.~B., Anderson S.~F., Strauss M.~A., York D.~G., Richards G.~T., Fan X., Knapp G.~R. et al., 2002, ApJS, 141, 267 
\bibitem[\protect\citeauthoryear{Hall}{2007}]{2007AJ.133.1271H} Hall P.~B., 2007, AJ, 133, 1271
\bibitem[\protect\citeauthoryear{Hamann et al.}{2013}]{2013MNRAS.435.133H} Hamann F., Chartas G., McGraw S., Rodriguez Hidalgo P., Shields J., Capellupo D., Charlton J. et al., 2013, MNRAS, 435, 133
\bibitem[\protect\citeauthoryear{Hamann et al.}{2019}]{2019MNRAS.487.5041H} Hamann F., Tripp T.~M., Rupke D., Veilleux S., 2019, MNRAS, 487, 5041
\bibitem[\protect\citeauthoryear{Hazard et al.}{1987}]{1987ApJ.323.263H} Hazard C., McMahon R.~G., Webb J.~K., Morton D.~C., 1987, ApJ, 323, 263 
\bibitem[Hennawi et al.(2009)]{2009ApJ.693L.49H} Hennawi J.~F., Prochaska J.~X., Kollmeier J., Zheng Z.,\ 2009, \apjl, 693, L49
\bibitem[Hopkins et al.(2012)]{2012MNRAS.420.320H} Hopkins P.~F., Hayward C.~C., Narayanan D., Hernquist L.,\ 2012, \mnras, 420, 320
\bibitem[Hutchings et al.(2002)]{2002AJ.124.2543H} Hutchings J.~B., Crenshaw D.~M., Kraemer S.~B., Gabel J.~R., Kaiser M.~E., Weistrop D., Gull T.~R.,\ 2002, \aj, 124, 2543
\bibitem[Jenkins(1986)]{1986ApJ.304.739J} Jenkins E.~B.,\ 1986, \apj, 304, 739
\bibitem[\protect\citeauthoryear{Jiang et al.}{2016}]{2016ApJ.821.1J} Jiang P., Zhou H., Pan X., Jiang N., Shu X., Wang H., Gu Q. et al., 2016, ApJ, 821, 1
\bibitem[Johansson et al.(2000)]{Johansson00} Johansson S., Zethson T., Hartman H., Ekberg J.~O., Ishibashi K., Davidson K., Gull T.~R., 2000, A\&A, 361, 977
\bibitem[Korista \etal(1992)]{Korista92} Korista K.~T., Weymann R.~J., Morris S.~L. \etal,\ 1992, ApJ, 401, 529
\bibitem[\protect\citeauthoryear{Kwan \& Krolik}{1979}]{1979ApJ.233L.91K} Kwan J., Krolik J.~H., 1979, ApJL, 233, L91
\bibitem[\protect\citeauthoryear{Laloux \& Petitjean}{2021}]{2021MNRAS.502.3855L} Laloux B., Petitjean P., 2021, MNRAS, 502, 3855
\bibitem[\protect\citeauthoryear{Leighly, Dietrich, \& Barber}{2011}]{2011ApJ.728.94L} Leighly K.~M., Dietrich M., Barber S., 2011, ApJ, 728, 94
\bibitem[\protect\citeauthoryear{Leighly et al.}{2018}]{2018ApJ.866.7L} Leighly K.~M., Terndrup D.~M., Gallagher S.~C., Richards G.~T., Dietrich M., 2018, ApJ, 866, 7
\bibitem[\protect\citeauthoryear{Liu et al.}{2015}]{2015ApJS.217.11L} Liu W.-J., Zhou H., Ji T., Yuan W., Wang T.-G., Jian G., Shi X. et al., 2015, ApJS, 217, 11
\bibitem[Markwardt(2009)]{Markwardt09} Markwardt C.~B., 2009, in Bohlender D., Dowler P., Durand D. eds, ASP Conf. Ser. Vol. 411, Astronomical Data Analysis Software and Systems XVIII. Astron. Soc. Pac., San Francisco, p. 251
\bibitem[Mathews \& Ferland(1987)]{Mathews87} Mathews W.~G., Ferland G.~J., 1987, ApJ, 323, 456
\bibitem[Morton(1991)]{1991ApJS.77.119M} Morton D.~C.,\ 1991, \apjs, 77, 119
\bibitem[Morton(2003)]{2003ApJS.149.205M} Morton D.~C.,\ 2003, \apjs, 149, 205
\bibitem[\protect\citeauthoryear{Murray et al.}{1995}]{1995ApJ.451.498M} Murray N., Chiang J., Grossman S.~A., Voit G.~M., 1995, ApJ, 451, 498
\bibitem[\protect\citeauthoryear{Noterdaeme et al.}{2019}]{2019A&A.627A.32N} Noterdaeme P., Balashev S., Krogager J.-K., Srianand R., Fathivavsari H., Petitjean P., Ledoux C., 2019, A\&A, 627, A32
\bibitem[Noterdaeme et al.(2014)]{2014A&A.566A.24N} Noterdaeme P., Petitjean P., P{\^a}ris I., Cai Z., Finley H., Ge J., Pieri M.~M., York D.~G.,\ 2014, \aap, 566, A24
\bibitem[\protect\citeauthoryear{Osterbrock \& Ferland}{2006}]{2006agna.book.O} Osterbrock D.~E., Ferland G.~J., 2006, Astrophysics of Gaseous Nebulae and Active Galactic Nuclei. University Science Books, Sausalito, CA
\bibitem[Pan \etal(2017)]{2017ApJ.835.218P} Pan X., Zhou H., Ge J. et al.,\ 2017, \apj, 835, 218
\bibitem[Peterson(1997)]{1997bk.Peterson} Peterson B.~M., 1997, An Introduction to Active Galactic Nuclei. Cambridge Univ. Press, Cambridge, 1997
\bibitem[\protect\citeauthoryear{Popovi{\'c} et al.}{2009}]{2009AJ.137.3548P} Popovi{\'c} L. {\v{C}}., Smirnova A.~A., Kova{\v{c}}evi{\'c} J., Moiseev A.~V., Afanasiev V.~L., 2009, AJ, 137, 3548
\bibitem[\protect\citeauthoryear{Sarkar et al.}{2021}]{2021ApJ.907.12S} Sarkar A., Ferland G.~J., Chatzikos M., Guzm{\'a}n F., van Hoof P.~A.~M., Smyth R.~T., Ramsbottom C.~A., et al., 2021, ApJ, 907, 12
\bibitem[\protect\citeauthoryear{Scargle, Caroff, \& Noerdlinger}{1970}]{1970ApJ.161L.115S} Scargle J.~D., Caroff L.~J., Noerdlinger P.~D., 1970, ApJL, 161, L115
\bibitem[Schlegel \etal(1998)]{Schlegel98} Schlegel D.~J., Finkbeiner D.~P., Davis M., 1998, ApJ, 500, 525
\bibitem[\protect\citeauthoryear{Schulze et al.}{2018}]{2018ApJ.853.167S} Schulze A., Misawa T., Zuo W., Wu X.-B., 2018, ApJ, 853, 167
\bibitem[Shapley et al.(2003)]{2003ApJ.588.65S} Shapley A.~E., Steidel C.~C., Pettini M., Adelberger K.~L.,\ 2003, \apj, 588, 65
\bibitem[\protect\citeauthoryear{Shi et al.}{2016a}]{2016ApJ.829.96S} Shi X.-H., Jiang P., Wang H.-Y., Zhang S.-H., Ji T., Liu W.-J., Zhou H.-Y., 2016a, ApJ, 829, 96
\bibitem[Shi et al.(2016b)]{2016ApJ.819.99S} Shi X., Zhou H., Shu X. et al.,\ 2016b, \apj, 819, 99
\bibitem[\protect\citeauthoryear{Smyth et al.}{2019}]{2019MNRAS.483..654S} Smyth R.~T., Ramsbottom C.~A., Keenan F.~P., Ferland G.~J., Ballance C.~P., 2019, MNRAS, 483, 654
\bibitem[\protect\citeauthoryear{Tayal \& Zatsarinny}{2018}]{2018PhRvA..98a2706T} Tayal S.~S., Zatsarinny O., 2018, PhRvA, 98, 012706. 
\bibitem[Tian \etal(2021)]{2021ApJ.914.13T} Tian Q., Shi X., Hao L. et al.,\ 2021, \apj, 914914.13, 13
\bibitem[\protect\citeauthoryear{Tombesi et al.}{2010}]{2010A&A.521A.57T} Tombesi F., Cappi M., Reeves J.~N., Palumbo G.~G.~C., Yaqoob T., Braito V., Dadina M., 2010, A\&A, 521, A57
\bibitem[\protect\citeauthoryear{Trump et al.}{2006}]{2006ApJS.165.1T} Trump J.~R., Hall P.~B., Reichard T.~A., Richards G.~T., Schneider D.~P., Vanden Berk D.~E., Knapp G.~R. et al., 2006, ApJS, 165, 1
\bibitem[Tyson(1988)]{1988ApJ.329L.57T} Tyson N.~D.,\ 1988, \apjl, 329, L57
\bibitem[Vanden Berk \etal(2001)]{Vanden01} Vanden Berk D.~E., Richards G.~T., Bauer A. \etal,\ 2001, AJ, 122, 549
\bibitem[Verner et al.(1999)]{Verner99} Verner E.~M., Verner D.~A., Korista K.~T., Ferguson J.~W., Hamann F., Ferland G.~J., 1999, ApJS, 120, 101
\bibitem[Vreeswijk et al.(2004)]{2004A&A.419.927V} Vreeswijk P.~M., Ellison S.~L., Ledoux C. et al.,\ 2004, \aap, 419, 927
\bibitem[Wada(2012)]{2012ApJ.758.66W} Wada K.,\ 2012, \apj, 758, 66
\bibitem[Watson et al.(2006)]{2006ApJ.652.1011W} Watson D., Fynbo J.~P.~U., Ledoux C. et al.,\ 2006, \apj, 652, 1011
\bibitem[\protect\citeauthoryear{Weymann, Carswell, \& Smith}{1981}]{1981ARA&A.19.41W} Weymann R.~J., Carswell R.~F., Smith M.~G., 1981, ARA\&A, 19, 41 
\bibitem[Wolfe et al.(1986)]{1986ApJS.61.249W} Wolfe A.~M., Turnshek D.~A., Smith H.~E., Cohen R.~D.,\ 1986, \apjs, 61, 249
\bibitem[\protect\citeauthoryear{Xie et al.}{2018}]{2018ApJ.858.32X} Xie X., Zhou H., Pan X., Jiang P., Shi X., Ji T., Zhang S. et al., 2018, ApJ, 858, 32
\bibitem[\protect\citeauthoryear{Zhang et al.}{2015}]{2015ApJ.815.113Z} Zhang S., Zhou H., Shi X., Shu X., Liu W., Ji T., Jiang P. et al., 2015, ApJ, 815, 113
\bibitem[\protect\citeauthoryear{Zsarg{\'o} \& Federman}{1998}]{1998ApJ.498.256Z} Zsarg{\'o} J., Federman S.~R., 1998, ApJ, 498, 256
\end{thebibliography}




\appendix
\onecolumn

\section{A comparison of atomic data for selected lines of \feii\ for four datasets.}
\label{comparison}

In this study, we have used the oscillator strengths (f-values) for a number of \feii\ lines for four datasets from \citet{Verner99}, \citet{2019MNRAS.483..654S}, \citet{2018PhRvA..98a2706T} and CHIANTI (Version 10.1) that are in the release of CLOUDY. Comparisons among the four datasets reveal large discrepancies in transition probabilities or oscillator strengths for part of transitions. The adoption of the most reliable f-values (and atomic data in general) can often significantly affect the photoionization modeling results. 
There is still a pressing need for improvements in highly reliable sets of atomic data for \feii.

We also observed a discrepancy in the total angular momentum (J) value for the same level between \citet{2018PhRvA..98a2706T} and the other three datasets, which consequently impacts the deduced oscillator strength of the corresponding transitions. The results are listed in Table\,\ref{tbl-5}. The index of the level ($\geq$ 114) contained in the Verner99 term, minus the number in parentheses, represents the modified level number that aligns with the atomic data for \feii\ in \citet{Verner99}.

Table\,\ref{tbl-4} list a comparison of atomic data for selected lines of \feii\ from four datasets. 
Wavelengths, index of the lower (l) and upper levels (u), and statistical weights (i.e. degeneracy) of lower and upper levels from \citet{Verner99} are given in columns 1 to 5, respectively. Transition probabilities (Einstein A coefficient) and then the deduced oscillator strengths (f-values) are included for four datasets, if they exist. The data denoted by * in the list were used to replace the counterpart of \citet{Verner99} to produce synthetic \feii\ spectra.The Einstein A coefficient for spontaneous decay is related to the absorption oscillator strength of the upward transition by
 
\begin{eqnarray}
A_{\rm ul} = \frac{8\pi^2e^2\nu^2}{m_ec^3}\frac{g_l}{g_u}f_{lu} = \frac{8\pi^2e^2}{m_ec\lambda^2}\frac{g_l}{g_u}f_{lu}, 
\end{eqnarray}
$m_e$ is electron mass.

\begin{table*}
  \centering
  \scriptsize
  \caption{Comparison of the disparity of the total angular momentum (J) value of the same level between \citet{2018PhRvA..98a2706T} and other three datasets. Energies are the data of four \feii\ data sets that are incorporated into the CLOUDY modelling code.} 
  \label{tbl-5}
	  \resizebox{1.\columnwidth}{!}{
  \begin{tabular}{lcccccccccccccccc} 
   \hline
   \raisebox{-1ex} {Configuration} & \raisebox{-1ex} {Term} & \multicolumn{3}{c}{Tayal18} && \multicolumn{3}{c}{Smyth19} && \multicolumn{3}{c}{Verner99} && \multicolumn{3}{c}{CHIANTI} \\[-.5ex] 
    \cline{3-5}\cline{7-9}\cline{11-13}\cline{15-17}  
    &  & Index & J & Energy$\ $(cm$^{-1})$ && Index & J & Energy$\ $(cm$^{-1})$ && Index & J & Energy$\ $(cm$^{-1})$ && Index & J & Energy$\ $(cm$^{-1})$  \\
    \hline 
    $3d^6(^3D)4s$ & $^4D$ & 49 & $1/2$ & 31364.5 && 49 & $3/2$ & 31364.4457 && 49 & $3/2$ & 31364.440 && 34 & $3/2$ & 31364.439   \\ 
                  &       & 50 & $3/2$ & 31368.5 && 50 & $1/2$ & 31368.4511 && 50 & $1/2$ & 31368.450 && 35 & $1/2$ & 31368.449   \\  
    $3d^6(^1F)4s$ & $^2F$ & 82 & $5/2$ & 44915.1 && 82 & $7/2$ & 44915.0489 && 82 & $7/2$ & 44915.046 &&  &  &    \\    
                  &       & 83 & $7/2$ & 44929.5 && 83 & $5/2$ & 44929.5233 && 83 & $5/2$ & 44929.550 &&  &  &    \\   
    $3d^6(^3F)4s$ & $^4F$ & 97 & $7/2$ & 50157.5 && 97 & $9/2$ & 50157.4631 && 97 & $9/2$ & 50157.452 && 61 & $9/2$ & 50157.453   \\    
                  &       & 98 & $9/2$ & 50187.8 && 98 & $7/2$ & 50187.8164 && 98 & $7/2$ & 50187.813 && 62 & $7/2$ & 50187.812   \\ 
    $3d^6(^3G)4p$ & $^4G$ & 159 & $5/2$ & 64040.9 && 157 & $7/2$ & 64040.8776 && 159(1)         & $7/2$ & 64040.886 && 98 & $7/2$ & 64040.887   \\    
                  &       & 160 & $7/2$ & 64087.4 && 158 & $5/2$ & 64087.3953 && 160(1)         & $5/2$ & 64087.418 && 99 & $5/2$ & 64087.418   \\ 
    $3d^6(^3G)4p$ & $^4H$ & 179 & $7/2$ & 66589.0 && 177 & $9/2$ & 66589.0326 && 179(1) & $9/2$ & 66589.008 && 110 & $9/2$ & 66589.008   \\    
                  &       & 181 & $9/2$ & 66672.3 && 179 & $7/2$ & 66672.3232 && 181(1) & $7/2$ & 66672.334 && 111 & $7/2$ & 66672.336   \\ 
    $3d^6(^1G)4p$ & $^2H$ & 184 & $9/2$ & 67516.3 && 182 & $11/2$ & 67516.3237 && 184(1) & $11/2$ & 67516.332 &&  &  &    \\    
                  &       & 185 & $11/2$ & 68000.8 && 186 & $9/2$ & 68000.7700 && 185(1) & $9/2$ & 68000.788 &&  &  &    \\ 
    $3d^54s^2$ & $^2G$ & 242 & $9/2$ & 78185.0 && 239 & $7/2$ & 78184.9950 && 232(2) & $7/2$ & 78185.030 &&  &  &    \\    
               &       & 248 & $7/2$ & 78577.3 && 243 & $9/2$ & 78577.2510 && 236(2) & $9/2$ & 78577.280 &&  &  &    \\                    
    $3d^6(^3D)4p$ & $^2P$ & 252 & $3/2$ & 78841.9 && 246 & $1/2$ & 78841.8606 && 239(2) & $1/2$ & 78841.960 &&  &  &    \\    
                  &       & 254 & $1/2$ & 79243.5 && 248 & $3/2$ & 79243.4552 && 241(2) & $3/2$ & 79243.600 &&  &  &    \\ 
    $3d^6(^3F)4p$ & $^4D$ & 310 & $7/2$ & 92899.3 &&     &       &            && 366(2) & $5/2$ & 92899.200 &&  &  &    \\    
                  &       & 311 & $5/2$ & 93129.9 && 393 & $7/2$ & 93129.8655 && 367(2) & $7/2$ & 93129.900 &&  &  &    \\ 
    \hline 
  \end{tabular}}
\end{table*} 

\begin{table*}
  \centering
  \scriptsize
  \caption{Comparison of atomic data for selected lines of \feii\ for four datasets. The \feii\ data sets of \citet{Verner99}, \citet{2019MNRAS.483..654S}, \citet{2018PhRvA..98a2706T} and CHIANTI that are in the release of CLOUDY.}
  \label{tbl-4}
  \resizebox{0.95\columnwidth}{!}{
  \begin{tabular}{lcccccccccccccccccc}
    \hline
   \raisebox{-1ex} {Wavelength} & \raisebox{-1ex} {$I_{\rm l}$} & \raisebox{-1ex} {$I_{\rm u}$} & \raisebox{-1ex} {$g_{\rm l}$} & \raisebox{-1ex} {$g_{\rm u}$} & \multicolumn{2}{c}{Verner99} && \multicolumn{3}{c}{Smyth19} && \multicolumn{3}{c}{Tayal18} && \multicolumn{3}{c}{CHIANTI} \\[-.5ex] 
    \cline{6-7}\cline{9-11}\cline{13-15}\cline{17-19}  
     &  &  &  &  & $A_{\rm ul}$ & $f_{\rm v}$ && $A_{\rm ul}$ & $f_{\rm s}$ & $f_{\rm s}/f_{\rm v}$ && $A_{\rm ul}$ & $f_{\rm t}$ & $f_{\rm t}/f_{\rm v}$ && $A_{\rm ul}$ & $f_{\rm c}$ & $f_{\rm c}/f_{\rm v}$ \\ [-.5ex]  
    \hline  
    2600.1729 & 1 &  64 & 10 & 10 & 2.39e+08 & 2.42e-01 && 2.53e+08 & 2.56e-01 & \textbf{1.06e+00} && 2.75e+08 & 2.79e-01 & \textbf{1.15e+00} && 2.37e+08 & 2.40e-01 & \textbf{9.92e-01}  \\ [-.1ex]
    2586.6499 & 1 &  65 & 10 &  8 & 8.84e+07 & 7.09e-02 && 8.76e+07 & 7.03e-02 & \textbf{9.92e-01} && 1.00e+08 & 8.02e-02 & \textbf{1.13e+00} && 6.83e+07 & 5.48e-02 & \textbf{7.73e-01}   \\ [-.1ex]
    2382.7651 & 1 &  69 & 10 & 12 & 3.36e+08 & 3.43e-01 && 3.40e+08 & 3.47e-01 & \textbf{1.01e+00} && 3.88e+08 & 3.96e-01 & \textbf{1.15e+00} && 3.35e+08 & 3.42e-01 & \textbf{9.97e-01}   \\ [-.1ex]
    2374.4612 & 1 &  70 & 10 & 10 & 3.90e+07 & 3.30e-02 && 4.69e+07 & 3.96e-02 & \textbf{1.20e+00} && 4.80e+07 & 4.06e-02 & \textbf{1.23e+00} && 6.24e+07 & 5.27e-02 & \textbf{1.60e+00}   \\ [-.1ex]
    2367.5906 & 1 &  71 & 10 &  8 & 3.21e+04 & 2.16e-05 && 8.41e+06 & 5.65e-03 & \textbf{2.62e+02} && 4.96e+05 & 3.33e-04 & \textbf{1.54e+01} && 6.32e+06 & 4.25e-03 & \textbf{1.97e+02}   \\ [-.1ex]
    2344.2139 & 1 &  75 & 10 &  8 & 1.90e+08 & 1.25e-01 && 1.79e+08 & 1.18e-01 & \textbf{9.44e-01} && 2.27e+08 & 1.50e-01 & \textbf{1.20e+00} && 1.92e+08 & 1.26e-01 & \textbf{1.01e+00}   \\ [-.1ex]
    2260.7805 & 1 &  78 & 10 & 10 & 3.42e+06 & 2.62e-03 && 1.41e+06 & \textbf{1.08e-03*} & \textbf{4.12e-01} && 3.02e+06 & 2.31e-03 & \textbf{8.82e-01} &&          &          & \textbf{        }   \\ [-.1ex]
    2234.4473 & 1 &  80 & 10 &  8 & 4.22e+04 & 2.53e-05 && 2.67e+05 & \textbf{1.60e-04*} & \textbf{6.32e+00} && 3.17e+04 & 1.90e-05 & \textbf{7.51e-01} &&          &          & \textbf{        }   \\ [-.5ex]  
    \hline 
    2612.6543 & 2 &  65 &  8 &  8 & 1.22e+08 & 1.25e-01 && 1.31e+08 & 1.34e-01 & \textbf{1.07e+00} && 1.41e+08 & 1.44e-01 & \textbf{1.15e+00} && 1.28e+08 & 1.31e-01 & \textbf{1.05e+00}   \\ [-.1ex]
    2599.1465 & 2 &  66 &  8 &  6 & 1.43e+08 & 1.09e-01 && 1.46e+08 & 1.11e-01 & \textbf{1.02e+00} && 1.63e+08 & 1.24e-01 & \textbf{1.14e+00} && 1.25e+08 & 9.48e-02 & \textbf{8.70e-01}   \\ [-.1ex]
    2396.3560 & 2 &  70 &  8 & 10 & 2.86e+08 & 3.08e-01 && 2.87e+08 & 3.09e-01 & \textbf{1.00e+00} && 3.29e+08 & 3.54e-01 & \textbf{1.15e+00} && 2.68e+08 & 2.89e-01 & \textbf{9.38e-01}   \\ [-.1ex]
    2389.3582 & 2 &  71 &  8 &  8 & 1.05e+08 & 8.99e-02 && 1.47e+08 & 1.26e-01 & \textbf{1.40e+00} && 1.19e+08 & 1.02e-01 & \textbf{1.13e+00} && 1.13e+08 & 9.68e-02 & \textbf{1.08e+00}   \\ [-.1ex]
    2365.5518 & 2 &  75 &  8 &  8 & 6.30e+07 & 5.29e-02 && 2.04e+07 & 1.71e-02 & \textbf{3.23e-01} && 7.79e+07 & 6.54e-02 & \textbf{1.24e+00} && 5.35e+07 & 4.49e-02 & \textbf{8.49e-01}   \\ [-.1ex]
    2333.5156 & 2 &  76 &  8 &  6 & 1.27e+08 & 7.78e-02 && 1.25e+08 & 7.65e-02 & \textbf{9.83e-01} && 1.51e+08 & 9.25e-02 & \textbf{1.19e+00} && 1.33e+08 & 8.14e-02 & \textbf{1.05e+00}   \\ [-.1ex]
    2280.6201 & 2 &  78 &  8 & 10 & 5.26e+06 & 5.13e-03 && 7.90e+05 & 7.70e-04 & \textbf{1.50e-01} && 4.61e+06 & \textbf{4.49e-03*} & \textbf{8.75e-01} &&          &          & \textbf{        }   \\ [-.1ex]
    2253.8254 & 2 &  80 &  8 &  8 & 4.23e+06 & 3.22e-03 && 7.03e+05 & 5.35e-04 & \textbf{1.66e-01} && 3.71e+06 & \textbf{2.83e-03*} & \textbf{8.79e-01} &&          &          & \textbf{        }   \\  [-.5ex]  
    \hline 
    2632.1082 & 3 &  65 &  6 &  8 & 5.95e+07 & 8.24e-02 && 7.31e+07 & 1.01e-01 & \textbf{1.23e+00} && 7.19e+07 & 9.96e-02 & \textbf{1.21e+00} && 9.01e+07 & 1.25e-01 & \textbf{1.52e+00}   \\ [-.1ex]
    2607.8665 & 3 &  67 &  6 &  4 & 1.81e+08 & 1.23e-01 && 1.90e+08 & 1.29e-01 & \textbf{1.05e+00} && 2.08e+08 & 1.41e-01 & \textbf{1.15e+00} && 1.73e+08 & 1.18e-01 & \textbf{9.59e-01}   \\ [-.1ex]
    2405.6187 & 3 &  71 &  6 &  8 & 2.17e+08 & 2.51e-01 && 1.67e+08 & 1.93e-01 & \textbf{7.69e-01} && 2.54e+08 & 2.94e-01 & \textbf{1.17e+00} && 2.10e+08 & 2.43e-01 & \textbf{9.68e-01}   \\ [-.1ex]
    2399.9729 & 3 &  72 &  6 &  6 & 1.45e+08 & 1.25e-01 && 1.52e+08 & 1.31e-01 & \textbf{1.05e+00} && 1.67e+08 & 1.44e-01 & \textbf{1.15e+00} && 1.52e+08 & 1.31e-01 & \textbf{1.05e+00}   \\ [-.1ex]
    2381.4888 & 3 &  75 &  6 &  8 & 3.35e+07 & 3.80e-02 && 8.08e+07 & 9.16e-02 & \textbf{2.41e+00} && 3.21e+07 & 3.64e-02 & \textbf{9.58e-01} && 8.72e+06 & 9.88e-03 & \textbf{2.60e-01}   \\ [-.1ex]
    2349.0222 & 3 &  76 &  6 &  6 & 1.08e+08 & 8.93e-02 && 1.00e+08 & 8.27e-02 & \textbf{9.26e-01} && 1.29e+08 & 1.07e-01 & \textbf{1.20e+00} && 9.32e+07 & 7.71e-02 & \textbf{8.63e-01}   \\ [-.1ex]
    2284.1899 & 3 &  79 &  6 &  8 & 7.76e+04 & 8.09e-05 && 4.58e+05 & \textbf{4.78e-04*} & \textbf{5.91e+00} && 4.61e+03 & 4.81e-06 & \textbf{ 5.95e-02} &&          &          & \textbf{        }   \\ [-.1ex]
    2268.2878 & 3 &  80 &  6 &  8 & 3.90e+06 & 4.01e-03 && 3.70e+05 & 3.81e-04 & \textbf{9.50e-02} && 3.54e+06 & \textbf{3.64e-03*} & \textbf{9.08e-01} &&          &          & \textbf{        }   \\ [-.5ex]  
    \hline 
    2360.7207 & 6 &  78 & 10 & 10 & 2.74e+07 & 2.29e-02 && 7.45e+07 & 6.22e-02 & \textbf{2.72e+00} && 2.60e+07 & 2.17e-02 & \textbf{9.48e-01} && 4.74e+07 & 3.96e-02 & \textbf{1.73e+00}   \\ [-.1ex]
    2348.8342 & 6 &  79 & 10 &  8 & 5.52e+07 & 3.65e-02 && 2.17e+08 & 1.44e-01 & \textbf{3.95e+00} && 5.52e+07 & 3.65e-02 & \textbf{1.00e+00} && 1.30e+08 & 8.60e-02 & \textbf{2.36e+00}   \\ [-.1ex]
    2332.0227 & 6 &  80 & 10 &  8 & 2.74e+07 & 1.79e-02 && 4.03e+06 & 2.63e-03 & \textbf{1.47e-01} && 2.83e+07 & 1.85e-02 & \textbf{1.03e+00} && 5.56e+06 & 3.63e-03 & \textbf{2.03e-01}   \\ [-.1ex]
    1696.7943 & 6 & 122 & 10 & 10 & 1.74e+07 & 7.51e-03 && 3.16e+04 & 1.36e-05 & \textbf{1.81e-03} && 3.17e+07 & 1.37e-02 & \textbf{1.82e+00} && 1.33e+07 & \textbf{5.74e-03*} & \textbf{7.64e-01}   \\ [-.1ex]
    1694.4836 & 6 & 124 & 10 & 12 & 7.87e+05 & 4.07e-04 &&          &          & \textbf{        } && 6.99e+06 & \textbf{3.61e-03*} & \textbf{8.87e+00} &&          &          & \textbf{        }   \\ [-.1ex]
    1661.3240 & 6 & 139 & 10 &  8 & 1.25e+06 & 4.14e-04 && 2.94e+05 & 9.73e-05 & \textbf{2.35e-01} && 1.62e+06 & 5.36e-04 & \textbf{1.29e+00} && 4.97e+06 & \textbf{1.65e-03*} & \textbf{3.99e+00}   \\ [-.1ex]
    1637.3990 & 6 & 151 & 10 &  8 & 4.41e+07 & 1.42e-02 && 4.18e+08 & 1.34e-01 & \textbf{9.44e+00} && 6.23e+07 & 2.00e-02 & \textbf{1.41e+00} && 2.38e+07 & \textbf{7.65e-03*} & \textbf{5.39e-01}   \\ [-.1ex]
    1612.8057 & 6 & 156 & 10 & 12 & 6.57e+07 & 3.07e-02 && 2.50e+07 & 1.17e-02 & \textbf{3.81e-01} && 6.02e+05 & 2.82e-04 & \textbf{9.19e-03} && 2.13e+07 & \textbf{9.97e-03*} & \textbf{3.25e-01}   \\ [-.1ex]
    1602.2103 & 6 & 160 & 10 &  8 & 1.05e+06 & 3.23e-04 && 3.41e+07 & 1.05e-02 & \textbf{3.25e+01} && 5.78e+05 & 1.78e-04 & \textbf{5.51e-01} &&          &          & \textbf{        }   \\ [-.1ex]
    1569.6746 & 6 & 168 & 10 & 12 & 2.30e+07 & 1.02e-02 && 1.56e+08 & 6.91e-02 & \textbf{6.77e+00} && 3.82e+06 & 1.69e-03 & \textbf{1.66e-01} && 4.22e+07 & 1.87e-02 & \textbf{1.83e+00}   \\ [-.1ex]
    1566.8217 & 6 & 169 & 10 & 10 & 5.20e+07 & 1.91e-02 && 1.67e+07 & 6.15e-03 & \textbf{3.22e-01} && 4.89e+04 & 1.80e-05 & \textbf{9.42e-04} && 3.51e+06 & 1.29e-03 & \textbf{6.75e-02}   \\ [-.5ex]  
    \hline    
    2344.6780 & 7 &  85 & 8 &  6 & 2.70e+07 & 1.67e-02 && 9.67e+05 & 5.98e-04 & \textbf{3.58e-02} && 2.89e+07 & 1.79e-02 & \textbf{1.07e+00} && 9.42e+06 & 5.82e-03 & \textbf{3.49e-01}   \\ [-.1ex]
    1676.8557 & 7 & 139 & 8 &  8 & 6.82e+06 & 2.87e-03 && 2.14e+07 & 9.02e-03 & \textbf{3.14e+00} && 9.65e+06 & \textbf{4.07e-03*} & \textbf{1.42e+00} && 3.56e+07 & 1.50e-02 & \textbf{5.23e+00}   \\ [-.1ex]
    1676.3613 & 7 & 140 & 8 & 10 & 1.32e+06 & 6.95e-04 && 2.07e+08 & 1.09e-01 & \textbf{1.57e+02} && 4.47e+06 & \textbf{2.35e-03*} & \textbf{3.38e+00} &&          &          & \textbf{        }   \\ [-.1ex]
    1674.4399 & 7 & 142 & 8 &  6 & 1.41e+06 & 4.45e-04 && 1.49e+04 & 4.70e-06 & \textbf{1.06e-02} && 3.90e+05 & 1.23e-04 & \textbf{2.76e-01} && 8.37e+06 & \textbf{2.64e-03*} & \textbf{5.93e+00}   \\ [-.1ex]
    1674.2563 & 7 & 143 & 8 & 10 & 1.08e+07 & 5.67e-03 && 5.19e+06 & 2.73e-03 & \textbf{4.81e-01} && 2.37e+07 & 1.24e-02 & \textbf{2.19e+00} && 3.89e+06 & \textbf{2.04e-03*} & \textbf{3.60e-01}   \\ [-.1ex]
    1625.5222 & 7 & 157 & 8 & 10 & 5.01e+07 & 2.48e-02 && 2.12e+07 & 1.05e-02 & \textbf{4.23e-01} && 7.24e+05 & 3.59e-04 & \textbf{1.45e-02} && 1.91e+07 & \textbf{9.46e-03*} & \textbf{3.81e-01}   \\ [-.1ex]
    1602.5167 & 7 & 163 & 8 & 10 & 2.23e+06 & 1.07e-03 && 7.61e+06 & \textbf{3.66e-03*} & \textbf{3.42e+00} && 5.07e+05 & 2.44e-04 & \textbf{2.28e-01} &&          &          & \textbf{        }   \\ [-.1ex]
    1580.6293 & 7 & 169 & 8 & 10 & 5.79e+07 & 2.71e-02 && 1.33e+08 & 6.23e-02 & \textbf{2.30e+00} && 9.18e+06 & 4.30e-03 & \textbf{1.59e-01} && 3.78e+07 & \textbf{1.77e-02*} & \textbf{6.53e-01}   \\ [-.1ex]
    1572.7560 & 7 & 171 & 8 & 10 & 8.56e+05 & 3.97e-04 && 6.48e+07 & 3.00e-02 & \textbf{7.56e+01} && 2.05e+07 & \textbf{9.50e-03*} & \textbf{2.39e+01} && 2.16e+07 & 1.00e-02 & \textbf{2.52e+01}   \\ [-.1ex]
    1560.2521 & 7 & 177 & 8 &  6 & 1.96e+07 & 5.36e-03 && 5.06e+07 & 1.39e-02 & \textbf{2.59e+00} && 1.75e+07 & 4.79e-03 & \textbf{8.94e-01} && 4.76e+07 & \textbf{1.30e-02*} & \textbf{2.43e+00}   \\ [-.1ex]
    1430.1665 & 7 & 201 & 8 &  8 & 1.00e+06 & 3.07e-04 && 6.65e+06 & \textbf{2.04e-03*} & \textbf{6.64e+00} && 5.55e+06 & 1.70e-03 & \textbf{5.54e+00} && 1.73e+06 & 5.29e-04 & \textbf{1.72e+00}   \\ [-.1ex]
    1424.7167 & 7 & 204 & 8 &  6 & 1.97e+07 & 4.50e-03 && 4.90e+07 & 1.12e-02 & \textbf{2.49e+00} && 3.14e+07 & 7.17e-03 & \textbf{1.59e+00} && 3.11e+07 & \textbf{7.10e-03*} & \textbf{1.58e+00}   \\ [-.1ex]
    1185.7124 & 7 & 291 & 8 &  6 & 1.78e+07 & 2.81e-03 && 3.24e+07 & \textbf{5.12e-03*} & \textbf{1.82e+00} && 8.40e+06 & 1.33e-03 & \textbf{4.73e-01} && 3.12e+07 & 4.93e-03 & \textbf{1.75e+00}   \\ [-.5ex]  
    \hline    
    1720.6127 &  8 & 125 &  6 &  8 & 5.88e+07 & 3.48e-02 && 6.92e+03 & 4.10e-06 & \textbf{1.18e-04} && 7.21e+07 & 4.27e-02 & \textbf{1.23e+00} && 1.32e+08 & 7.82e-02 & \textbf{2.25e+00}   \\ [-.1ex]
    1718.1010 &  8 & 128 &  6 &  6 & 1.41e+07 & 6.24e-03 && 3.77e+07 & 1.67e-02 & \textbf{2.68e+00} && 4.65e+07 & 2.06e-02 & \textbf{3.30e+00} && 2.13e+07 & 9.43e-03 & \textbf{1.51e+00}   \\ [-.1ex] 
    1698.1353 &  8 & 136 &  6 &  8 & 1.39e+06 & 8.01e-04 && 7.71e+05 & 4.44e-04 & \textbf{5.54e-01} && 3.68e+06 & \textbf{2.12e-03*} & \textbf{2.65e+00} && 5.74e+05 & 3.31e-04 & \textbf{4.13e-01}   \\ [-.1ex] 
    1688.4030 &  8 & 139 &  6 &  8 & 3.67e+06 & 2.09e-03 && 6.00e+06 & 3.42e-03 & \textbf{1.64e+00} && 2.00e+07 & 1.14e-02 & \textbf{5.45e+00} && 6.13e+06 & \textbf{3.49e-03*} & \textbf{1.67e+00}   \\ [-.1ex] 
    1683.3156 &  8 & 145 &  6 &  4 & 2.69e+05 & 7.62e-05 && 2.76e+06 & 7.82e-04 & \textbf{1.03e+01} && 8.41e+06 & 2.38e-03 & \textbf{3.12e+01} && 9.23e+06 & \textbf{2.61e-03*} & \textbf{3.43e+01}   \\ [-.1ex] 
    1681.1107 &  8 & 147 &  6 &  8 & 6.47e+06 & 3.66e-03 && 1.88e+08 & 1.06e-01 & \textbf{2.90e+01} && 7.33e+06 & \textbf{4.14e-03*} & \textbf{1.13e+00} &&          &          & \textbf{        }   \\ [-.1ex] 
    1654.6696 &  8 & 153 &  6 &  6 & 9.00e+05 & 3.69e-04 && 4.48e+07 & 1.84e-02 & \textbf{4.99e+01} && 2.22e+06 & 9.11e-04 & \textbf{2.47e+00} && 4.50e+06 & \textbf{1.85e-03*} & \textbf{5.01e+00}   \\ [-.1ex] 
    1584.9521 &  8 & 170 &  6 &  8 & 5.80e+07 & 2.91e-02 && 1.34e+08 & 6.73e-02 & \textbf{2.31e+00} && 9.50e+06 & 4.77e-03 & \textbf{1.64e-01} && 3.52e+07 & \textbf{1.77e-02*} & \textbf{6.08e-01}   \\ [-.1ex] 
    1444.9807 &  8 & 195 &  6 &  4 & 3.59e+06 & 7.49e-04 && 9.80e+06 & \textbf{2.05e-03*} & \textbf{2.74e+00} && 3.87e+05 & 8.08e-05 & \textbf{1.08e-01} &&          &          & \textbf{        }   \\ [-.1ex] 
    1440.9105 &  8 & 199 &  6 &  6 & 7.34e+05 & 2.28e-04 && 5.92e+06 & \textbf{1.84e-03*} & \textbf{8.07e+00} && 5.39e+06 & 1.68e-03 & \textbf{7.37e+00} && 1.52e+06 & 4.74e-04 & \textbf{2.08e+00}   \\ [-.1ex] 
    1434.9958 &  8 & 203 &  6 &  4 & 1.89e+07 & 3.89e-03 && 4.99e+07 & \textbf{1.03e-02*} & \textbf{2.65e+00} && 3.30e+07 & 6.79e-03 & \textbf{1.75e+00} && 3.00e+07 & 6.17e-03 & \textbf{1.59e+00}   \\ [-.1ex] 
    1433.0438 &  8 & 204 &  6 &  6 & 2.85e+06 & 8.77e-04 && 6.75e+06 & \textbf{2.08e-03*} & \textbf{2.37e+00} && 3.56e+06 & 1.10e-03 & \textbf{1.25e+00} && 6.51e+06 & 2.00e-03 & \textbf{2.28e+00}   \\ [-.5ex]  
    \hline 
    1726.3916 &  9 & 128 &  4 &  6 & 3.23e+07 & 2.16e-02 && 1.74e+08 & 1.17e-01 & \textbf{5.42e+00} && 1.29e+08 & 8.65e-02 & \textbf{4.00e+00} && 1.30e+08 & 8.73e-02 & \textbf{4.04e+00}   \\ [-.1ex] 
    1691.2729 &  9 & 145 &  4 &  4 & 2.56e+07 & 1.10e-02 && 1.85e+07 & 7.93e-03 & \textbf{7.21e-01} && 4.13e+07 & 1.77e-02 & \textbf{1.61e+00} && 3.64e+07 & \textbf{1.56e-02*} & \textbf{1.42e+00}   \\ [-.1ex] 
    1674.7161 &  9 & 150 &  4 &  2 & 9.12e+07 & 1.92e-02 && 7.19e+06 & 1.51e-03 & \textbf{7.86e-02} && 4.74e+07 & 9.97e-03 & \textbf{5.19e-01} && 1.16e+08 & 2.43e-02 & \textbf{1.27e+00}   \\ [-.1ex] 
    1654.4779 &  9 & 155 &  4 &  2 & 7.15e+07 & 1.47e-02 && 5.06e+08 & 1.04e-01 & \textbf{7.07e+00} && 1.24e+08 & \textbf{2.54e-02*} & \textbf{1.73e+00} && 2.59e+07 & 5.31e-03 & \textbf{3.61e-01}   \\ [-.1ex] 
    1640.1521 &  9 & 159 &  4 &  6 & 5.90e+07 & 3.57e-02 && 1.92e+07 & 1.16e-02 & \textbf{3.25e-01} &&          &          & \textbf{        } && 1.74e+07 & \textbf{1.05e-02*} & \textbf{2.94e-01}   \\ [-.1ex] 
    1588.2897 &  9 & 172 &  4 &  6 & 4.82e+07 & 2.73e-02 && 1.31e+08 & 7.43e-02 & \textbf{2.72e+00} && 1.33e+07 & 7.55e-03 & \textbf{2.77e-01} && 3.49e+07 & \textbf{1.98e-02*} & \textbf{7.25e-01}   \\ [-.1ex] 
    1577.1666 &  9 & 177 &  4 &  6 & 1.69e+07 & 9.45e-03 && 9.16e+07 & 5.12e-02 & \textbf{5.42e+00} && 2.86e+07 & 1.60e-02 & \textbf{1.69e+00} && 3.44e+07 & \textbf{1.93e-02*} & \textbf{2.04e+00}   \\ [-.1ex] 
    1574.9222 &  9 & 179 &  4 &  4 & 1.21e+08 & 4.50e-02 && 4.47e+08 & 1.66e-01 & \textbf{3.69e+00} && 1.23e+08 & 4.57e-02 & \textbf{1.02e+00} && 2.08e+08 & 7.72e-02 & \textbf{1.72e+00}   \\ [-.1ex] 
    1448.1937 &  9 & 197 &  4 &  4 & 1.15e+05 & 3.62e-05 && 5.31e+06 & \textbf{1.67e-03*} & \textbf{4.61e+01} && 5.09e+06 & 1.60e-03 & \textbf{4.42e+01} && 1.74e+06 & 5.49e-04 & \textbf{1.52e+01}   \\ [-.1ex] 
    1442.7465 &  9 & 202 &  4 &  2 & 1.81e+07 & 2.82e-03 && 4.00e+07 & 6.24e-03 & \textbf{2.21e+00} && 4.07e+07 & \textbf{6.35e-03*} & \textbf{2.25e+00} && 3.67e+07 & 5.73e-03 & \textbf{2.03e+00}   \\ [-.1ex] 
    1440.7747 &  9 & 203 &  4 &  4 & 3.75e+06 & 1.17e-03 && 9.84e+06 & \textbf{3.06e-03*} & \textbf{2.62e+00} && 5.35e+06 & 1.66e-03 & \textbf{1.42e+00} && 7.37e+06 & 2.29e-03 & \textbf{1.96e+00}   \\ [-.5ex]  
    \hline  
    2717.5020 & 10 &  80 &  8 &  8 & 7.35e+04 & 8.14e-05 && 1.06e+08 & 1.17e-01 & \textbf{1.44e+03} && 3.09e+05 & 3.42e-04 & \textbf{4.20e+00} && 3.51e+07 & 3.89e-02 & \textbf{4.78e+02}   \\ [-.1ex] 
    2715.2175 & 10 &  81 &  8 &  6 & 5.48e+07 & 4.54e-02 && 4.48e+07 & 3.71e-02 & \textbf{8.17e-01} && 6.74e+07 & 5.59e-02 & \textbf{1.23e+00} && 4.88e+07 & 4.04e-02 & \textbf{8.90e-01}   \\ [-.1ex] 
    2693.6333 & 10 &  85 &  8 &  6 & 1.59e+06 & 1.30e-03 && 1.59e+07 & 1.30e-02 & \textbf{1.00e+01} && 2.85e+06 & \textbf{2.33e-03*} & \textbf{1.79e+00} && 2.44e+06 & 1.99e-03 & \textbf{1.53e+00}   \\ [-.1ex] 
    2563.3042 & 10 &  88 &  8 &  6 & 1.83e+08 & 1.35e-01 && 1.95e+08 & 1.44e-01 & \textbf{1.07e+00} && 2.05e+08 & 1.51e-01 & \textbf{1.12e+00} && 1.93e+08 & 1.43e-01 & \textbf{1.06e+00}   \\ [-.1ex] 
    1722.4318 & 10 & 171 &  8 & 10 & 3.52e+06 & 1.96e-03 && 1.68e+07 & \textbf{9.34e-03*} & \textbf{4.77e+00} && 7.31e+06 & 4.06e-03 & \textbf{2.07e+00} && 7.33e+06 & 4.07e-03 & \textbf{2.08e+00}   \\ [-.1ex] 
    1183.8287 & 10 & 358 &  8 & 10 & 4.88e+07 & 1.28e-02 && 1.06e+08 & \textbf{2.78e-02*} & \textbf{2.17e+00} && 3.68e+07 & 9.66e-03 & \textbf{7.55e-01} &&          &          & \textbf{        }   \\ [-.1ex] 
    1169.1902 & 10 & 368 &  8 & 10 & 3.91e+07 & 1.00e-02 && 5.57e+07 & \textbf{1.43e-02*} & \textbf{1.43e+00} && 3.11e+07 & 7.97e-03 & \textbf{7.97e-01} && 2.24e+05 & 5.73e-05 & \textbf{5.73e-03}   \\  [-.5ex]  
    \hline 
    2773.5447 & 11 &  79 &  6 &  8 & 1.07e+05 & 1.65e-04 && 1.16e+08 & 1.78e-01 & \textbf{1.08e+03} && 3.08e+05 & 4.74e-04 & \textbf{2.87e+00} && 3.43e+07 & 5.28e-02 & \textbf{3.20e+02}   \\ [-.1ex] 
    2728.3469 & 11 &  84 &  6 &  4 & 9.26e+07 & 6.89e-02 && 7.13e+07 & \textbf{5.30e-02*} & \textbf{7.69e-01} && 1.14e+08 & 8.48e-02 & \textbf{1.23e+00} && 8.86e+07 & 6.59e-02 & \textbf{9.56e-01}   \\ [-.1ex] 
    2725.6914 & 11 &  85 &  6 &  6 & 9.37e+06 & 1.04e-02 && 1.11e+08 & 1.24e-01 & \textbf{1.19e+01} && 8.60e+06 & 9.58e-03 & \textbf{9.21e-01} && 5.94e+07 & 6.61e-02 & \textbf{6.36e+00}   \\ [-.1ex] 
    1874.9717 & 11 & 136 &  6 &  8 & 2.13e+06 & 1.50e-03 && 7.81e+06 & \textbf{5.49e-03*} & \textbf{3.66e+00} && 1.91e+06 & 1.34e-03 & \textbf{8.93e-01} && 1.15e+06 & 8.05e-04 & \textbf{5.37e-01}   \\ [-.1ex] 
    1833.0764 & 11 & 151 &  6 &  8 & 2.23e+06 & 1.50e-03 && 1.71e+07 & 1.15e-02 & \textbf{7.67e+00} && 3.62e+06 & \textbf{2.43e-03*} & \textbf{1.62e+00} && 1.35e+06 & 9.04e-04 & \textbf{6.03e-01}   \\ [-.1ex] 
    1822.1230 & 11 & 153 &  6 &  6 & 4.26e+06 & 2.12e-03 && 9.84e+06 & \textbf{4.90e-03*} & \textbf{2.31e+00} && 7.62e+06 & 3.79e-03 & \textbf{1.79e+00} && 5.51e+06 & 2.74e-03 & \textbf{1.29e+00}   \\ [-.1ex] 
    1273.2739 & 11 & 292 &  6 &  8 & 1.18e+04 & 3.82e-06 && 2.21e+07 & \textbf{7.16e-03*} & \textbf{1.87e+03} && 3.98e+03 & 1.29e-06 & \textbf{3.38e-01} && 1.14e+05 & 3.70e-05 & \textbf{9.69e+00}   \\ 
    \hline
 \end{tabular}}
\end{table*}

\begin{table*}
  \centering
  \scriptsize
  \contcaption{}
  \resizebox{1.\columnwidth}{!}{
  \begin{tabular}{lcccccccccccccccccc} 
    \hline
   \raisebox{-1ex} {Wavelength} & \raisebox{-1ex} {$I_{\rm l}$} & \raisebox{-1ex} {$I_{\rm u}$} & \raisebox{-1ex} {$g_{\rm l}$} & \raisebox{-1ex} {$g_{\rm u}$} & \multicolumn{2}{c}{Verner99} && \multicolumn{3}{c}{Smyth19} && \multicolumn{3}{c}{Tayal18} && \multicolumn{3}{c}{CHIANTI} \\[-.5ex] 
    \cline{6-7}\cline{9-11}\cline{13-15}\cline{17-19}  
     &  &  &  &  & $A_{\rm ul}$ & $f_{\rm v}$ && $A_{\rm ul}$ & $f_{\rm s}$ & $f_{\rm s}/f_{\rm v}$ && $A_{\rm ul}$ & $f_{\rm t}$ & $f_{\rm t}/f_{\rm v}$ && $A_{\rm ul}$ & $f_{\rm c}$ & $f_{\rm c}/f_{\rm v}$ \\
    \hline 
    2769.7527 & 12 &  81 &  4 &  6 & 3.87e+06 & 6.68e-03 && 1.31e+08 & 2.26e-01 & \textbf{3.38e+01} && 2.95e+06 & 5.09e-03 & \textbf{7.62e-01} && 5.64e+07 & 9.74e-02 & \textbf{1.46e+01}   \\ 
    2731.5427 & 12 &  87 &  4 &  4 & 2.65e+07 & 2.96e-02 && 1.03e+08 & 1.15e-01 & \textbf{3.89e+00} && 2.74e+07 & 3.06e-02 & \textbf{1.03e+00} && 6.79e+07 & 7.60e-02 & \textbf{2.57e+00}   \\ 
    1842.2397 & 12 & 152 &  4 &  4 & 3.00e+06 & 1.53e-03 && 3.87e+04 & 1.97e-05 & \textbf{1.29e-02} && 1.37e+06 & 6.97e-04 & \textbf{4.56e-01} && 3.39e+06 & \textbf{1.73e-03*} & \textbf{1.13e+00}   \\ 
    1831.7527 & 12 & 153 &  4 &  6 & 3.07e+06 & 2.32e-03 && 8.63e+06 & \textbf{6.51e-03*} & \textbf{2.81e+00} && 6.18e+06 & 4.66e-03 & \textbf{2.01e+00} && 2.21e+06 & 1.67e-03 & \textbf{7.20e-01}   \\ 
    1825.3286 & 12 & 154 &  4 &  4 & 3.75e+06 & 1.87e-03 && 1.40e+07 & \textbf{6.99e-03*} & \textbf{3.74e+00} && 6.86e+06 & 3.43e-03 & \textbf{1.83e+00} && 3.83e+06 & 1.91e-03 & \textbf{1.02e+00}   \\ 
    1822.1895 & 12 & 155 &  4 &  2 & 6.25e+06 & 1.56e-03 && 2.18e+07 & \textbf{5.43e-03*} & \textbf{3.48e+00} && 1.10e+07 & 2.74e-03 & \textbf{1.76e+00} && 4.81e+06 & 1.20e-03 & \textbf{7.69e-01}   \\ 
    1728.8521 & 12 & 177 &  4 &  6 & 3.20e+06 & 2.15e-03 && 8.71e+06 & \textbf{5.85e-03*} & \textbf{2.72e+00} && 4.21e+06 & 2.83e-03 & \textbf{1.32e+00} && 5.44e+06 & 3.65e-03 & \textbf{1.70e+00}   \\ 
    1646.1847 & 12 & 187 &  4 &  2 & 1.38e+08 & 2.80e-02 && 2.40e+08 & \textbf{4.88e-02*} & \textbf{1.74e+00} && 1.22e+06 & 2.48e-04 & \textbf{8.86e-03} &&          &          & \textbf{        }   \\ 
    \hline 
    2762.6294 & 13 &  84 &  2 &  4 & 1.24e+07 & 2.84e-02 && 1.10e+08 & 2.52e-01 & \textbf{8.87e+00} && 1.12e+07 & 2.56e-02 & \textbf{9.01e-01} && 6.09e+07 & 1.39e-01 & \textbf{4.89e+00}   \\ 
    2594.5042 & 13 &  89 &  2 &  4 & 1.43e+07 & 2.89e-02 && 1.24e+07 & 2.50e-02 & \textbf{8.65e-01} && 1.66e+07 & 3.35e-02 & \textbf{1.16e+00} && 1.16e+07 & 2.35e-02 & \textbf{8.13e-01}   \\ 
    2578.6948 & 13 &  90 &  2 &  2 & 1.24e+08 & 1.24e-01 && 1.40e+08 & 1.40e-01 & \textbf{1.13e+00} && 1.41e+08 & 1.41e-01 & \textbf{1.14e+00} && 1.18e+08 & 1.18e-01 & \textbf{9.52e-01}   \\ 
    1827.7285 & 13 & 155 &  2 &  2 & 6.59e+06 & 3.30e-03 && 2.30e+07 & \textbf{1.15e-02*} & \textbf{3.48e+00} && 1.10e+07 & 5.51e-03 & \textbf{1.67e+00} && 4.76e+06 & 2.39e-03 & \textbf{7.24e-01}   \\ 
    1731.1255 & 13 & 179 &  2 &  4 & 3.27e+06 & 2.94e-03 && 9.40e+06 & \textbf{8.45e-03*} & \textbf{2.87e+00} && 4.21e+06 & 3.78e-03 & \textbf{1.29e+00} && 5.05e+06 & 4.54e-03 & \textbf{1.54e+00}   \\ 
    1578.1282 & 13 & 198 &  2 &  2 & 9.53e+06 & 3.56e-03 && 6.47e+06 & 2.42e-03 & \textbf{6.80e-01} && 5.04e+06 & 1.88e-03 & \textbf{5.28e-01} && 8.58e+06 & 3.20e-03 & \textbf{8.99e-01}   \\ 
    \hline
    2985.6956 & 14 &  88 &  6 &  6 & 2.91e+07 & 3.89e-02 && 1.09e+08 & 1.46e-01 & \textbf{3.75e+00} && 2.83e+07 & \textbf{3.78e-02*} & \textbf{9.72e-01} && 4.38e+07 & 5.85e-02 & \textbf{1.50e+00}   \\ 
    2948.5159 & 14 &  89 &  6 &  4 & 1.93e+07 & 1.68e-02 && 2.35e+07 & \textbf{2.04e-02*} & \textbf{1.21e+00} && 1.90e+07 & 1.65e-02 & \textbf{9.82e-01} && 2.92e+07 & 2.54e-02 & \textbf{1.51e+00}   \\ 
    2165.0154 & 14 & 115 &  6 &  4 & 1.14e+07 & 5.34e-03 && 2.48e+07 & 1.16e-02 & \textbf{2.17e+00} && 1.68e+07 & 7.87e-03 & \textbf{1.47e+00} && 6.59e+07 & 3.09e-02 & \textbf{5.79e+00}   \\ 
    2130.9270 & 14 & 118 &  6 &  6 & 3.87e+06 & 2.63e-03 && 9.48e+06 & \textbf{6.45e-03*} & \textbf{2.45e+00} && 2.43e+06 & 1.65e-03 & \textbf{6.27e-01} && 5.84e+07 & 3.98e-02 & \textbf{1.51e+01}   \\ 
    2031.8815 & 14 & 149 &  6 &  6 & 2.24e+04 & 1.39e-05 && 1.59e+07 & 9.84e-03 & \textbf{7.08e+02} && 4.54e+06 & 2.81e-03 & \textbf{2.02e+02} && 2.86e+07 & 1.77e-02 & \textbf{1.27e+03}   \\ 
    2021.4015 & 14 & 151 &  6 &  8 & 2.70e+07 & 2.21e-02 && 5.72e+06 & 4.67e-03 & \textbf{2.11e-01} && 2.93e+07 & 2.39e-02 & \textbf{1.08e+00} && 4.26e+07 & \textbf{3.48e-02*} & \textbf{1.57e+00}   \\ 
    1693.4757 & 14 & 203 &  6 &  4 & 1.97e+07 & 5.65e-03 && 4.75e+07 & \textbf{1.36e-02*} & \textbf{2.41e+00} && 1.98e+06 & 5.68e-04 & \textbf{1.01e-01} && 2.46e+06 & 7.05e-04 & \textbf{1.25e-01}   \\ 
    1690.7577 & 14 & 204 &  6 &  6 & 2.87e+07 & 1.23e-02 && 7.34e+07 & \textbf{3.15e-02*} & \textbf{2.56e+00} && 1.48e+07 & 6.34e-03 & \textbf{5.15e-01} && 1.48e+07 & 6.36e-03 & \textbf{5.17e-01}   \\ 
    1689.8324 & 14 & 206 &  6 &  8 & 1.98e+07 & 1.13e-02 && 4.46e+07 & 2.55e-02 & \textbf{2.26e+00} && 2.12e+07 & 1.21e-02 & \textbf{1.07e+00} && 4.95e+07 & \textbf{2.83e-02*} & \textbf{2.50e+00}   \\ 
    1364.3837 & 14 & 291 &  6 &  6 & 3.45e+07 & 9.63e-03 && 7.06e+07 & \textbf{1.97e-02*} & \textbf{2.05e+00} && 2.43e+07 & 6.78e-03 & \textbf{7.04e-01} && 3.18e+07 & 8.86e-03 & \textbf{9.20e-01}   \\ 
    1296.0840 & 14 & 327 &  6 &  4 & 2.97e+06 & 4.99e-04 && 3.27e+08 & \textbf{5.49e-02*} & \textbf{1.10e+02} && 1.80e+08 & 3.02e-02 & \textbf{6.05e+01} && 2.32e+08 & 3.89e-02 & \textbf{7.80e+01}   \\ 
    \hline
    3003.5208 & 15 &  88 &  4 &  6 & 1.06e+07 & 2.15e-02 && 3.39e+07 & 6.88e-02 & \textbf{3.20e+00} && 1.02e+07 & 2.07e-02 & \textbf{9.63e-01} && 1.84e+07 & \textbf{3.74e-02*} & \textbf{1.74e+00}   \\ 
    2965.8987 & 15 &  89 &  4 &  4 & 5.95e+06 & 7.85e-03 && 2.49e+07 & 3.28e-02 & \textbf{4.18e+00} && 5.87e+06 & 7.74e-03 & \textbf{9.86e-01} && 8.50e+06 & \textbf{1.12e-02*} & \textbf{1.43e+00}   \\ 
    2945.2573 & 15 &  90 &  4 &  2 & 3.54e+07 & 2.30e-02 && 8.46e+07 & 5.50e-02 & \textbf{2.39e+00} && 3.47e+07 & 2.26e-02 & \textbf{9.83e-01} && 5.43e+07 & \textbf{3.53e-02*} & \textbf{1.53e+00}   \\ 
    2174.3728 & 15 & 115 &  4 &  4 & 7.76e+05 & 5.50e-04 && 2.93e+06 & 2.08e-03 & \textbf{3.78e+00} && 5.70e+06 & \textbf{4.04e-03*} & \textbf{7.35e+00} && 4.34e+07 & 3.07e-02 & \textbf{5.58e+01}   \\
    2139.9915 & 15 & 118 &  4 &  6 & 1.43e+05 & 1.47e-04 && 5.17e+05 & 5.32e-04 & \textbf{3.62e+00} && 8.57e+05 & 8.83e-04 & \textbf{6.01e+00} && 2.47e+07 & \textbf{2.55e-02*} & \textbf{1.73e+02}   \\ 
    2111.3928 & 15 & 127 &  4 &  2 & 5.79e+06 & 1.93e-03 && 1.48e+07 & 4.95e-03 & \textbf{2.56e+00} && 3.32e+06 & 1.11e-03 & \textbf{5.75e-01} && 7.15e+07 & \textbf{2.39e-02*} & \textbf{1.24e+01}   \\ 
    2016.1375 & 15 & 153 &  4 &  6 & 2.31e+07 & 2.11e-02 && 1.38e+06 & 1.26e-03 & \textbf{5.97e-02} && 2.24e+07 & 2.05e-02 & \textbf{9.72e-01} && 3.01e+07 & \textbf{2.75e-02*} & \textbf{1.30e+00}   \\ 
    2008.3578 & 15 & 154 &  4 &  4 & 1.63e+07 & 9.86e-03 && 1.02e+07 & 6.17e-03 & \textbf{6.26e-01} && 1.48e+07 & 8.95e-03 & \textbf{9.08e-01} && 2.32e+07 & \textbf{1.40e-02*} & \textbf{1.42e+00}   \\ 
    1715.5153 & 15 & 194 &  4 &  6 & 5.09e+07 & 3.37e-02 && 1.34e+08 & \textbf{8.87e-02*} & \textbf{2.63e+00} && 2.58e+07 & 1.71e-02 & \textbf{5.07e-01} && 3.47e+07 & 2.29e-02 & \textbf{6.80e-01}   \\ 
    1372.2920 & 15 & 287 &  4 &  4 & 6.40e+07 & 1.81e-02 && 1.27e+08 & \textbf{3.59e-02*} & \textbf{1.98e+00} && 4.00e+07 & 1.13e-02 & \textbf{6.24e-01} && 5.55e+07 & 1.57e-02 & \textbf{8.67e-01}   \\ 
    \hline
    2986.4158 & 16 &  89 &  2 &  4 & 1.60e+07 & 4.28e-02 && 4.92e+07 & 1.32e-01 & \textbf{3.08e+00} && 1.55e+07 & 4.14e-02 & \textbf{9.67e-01} && 2.60e+07 & 6.96e-02 & \textbf{1.63e+00}   \\ 
    2965.4890 & 16 &  90 &  2 &  2 & 6.54e+06 & 8.62e-03 && 1.50e+07 & 1.98e-02 & \textbf{2.30e+00} && 6.35e+06 & 8.37e-03 & \textbf{9.71e-01} && 1.06e+07 & \textbf{1.40e-02*} & \textbf{1.62e+00}   \\ 
    2185.3799 & 16 & 115 &  2 &  4 & 4.09e+04 & 5.86e-05 && 3.57e+05 & 5.11e-04 & \textbf{8.72e+00} && 2.06e+06 & 2.95e-03 & \textbf{5.03e+01} && 2.14e+07 & \textbf{3.06e-02*} & \textbf{5.22e+02}   \\ 
    2108.4619 & 16 & 131 &  2 &  4 & 1.45e+06 & 1.93e-03 && 4.37e+06 & 5.83e-03 & \textbf{3.02e+00} && 1.27e+06 & 1.69e-03 & \textbf{8.76e-01} && 3.59e+07 & \textbf{4.79e-02*} & \textbf{2.48e+01}   \\ 
    2043.9761 & 16 & 150 &  2 &  2 & 2.85e+06 & 1.79e-03 && 6.59e+07 & 4.13e-02 & \textbf{2.31e+01} && 1.44e+07 & \textbf{9.02e-03*} & \textbf{5.04e+00} && 7.81e+07 & 4.89e-02 & \textbf{2.73e+01}   \\ 
    2038.4292 & 16 & 152 &  2 &  4 & 4.93e+05 & 6.14e-04 && 3.25e+07 & 4.05e-02 & \textbf{6.60e+01} && 2.97e+06 & \textbf{3.70e-03*} & \textbf{6.03e+00} && 3.94e+07 & 4.90e-02 & \textbf{7.98e+01}   \\ 
    2017.7448 & 16 & 154 &  2 &  4 & 1.49e+07 & 1.82e-02 && 3.86e+06 & 4.71e-03 & \textbf{2.59e-01} && 1.20e+07 & 1.46e-02 & \textbf{8.02e-01} && 1.78e+07 & \textbf{2.18e-02*} & \textbf{1.20e+00}   \\ 
    2013.9095 & 16 & 155 &  2 &  2 & 2.75e+07 & 1.67e-02 && 1.63e+07 & 9.91e-03 & \textbf{5.93e-01} && 2.06e+07 & 1.25e-02 & \textbf{7.49e-01} && 3.59e+07 & \textbf{2.18e-02*} & \textbf{1.31e+00}   \\ 
    1379.6149 & 16 & 284 &  2 &  2 & 1.01e+08 & 2.88e-02 && 2.01e+08 & \textbf{5.74e-02*} & \textbf{1.99e+00} && 5.97e+07 & 1.70e-02 & \textbf{5.90e-01} && 8.53e+07 & 2.43e-02 & \textbf{8.44e-01}   \\ 
    1298.8018 & 16 & 332 &  2 &  4 & 3.78e+05 & 1.91e-04 && 7.56e+05 & 3.82e-04 & \textbf{2.00e+00} &&          &          & \textbf{        } && 3.04e+07 & \textbf{1.54e-02*} & \textbf{8.06e+01}   \\ 
    \hline
    1935.2960 & 17 & 183 & 10 & 12 & 8.38e+06 & 5.65e-03 && 9.50e+07 & 6.40e-02 & \textbf{1.13e+01} && 1.05e+04 & 5.90e-06 & \textbf{1.04e-03} &&          &          & \textbf{        }   \\ 
    1936.8048 & 18 & 184 &  8 & 10 & 6.76e+06 & 4.75e-03 && 8.93e+07 & \textbf{6.28e-02*} & \textbf{1.32e+01} &&          &          & \textbf{        } &&          &          & \textbf{        }   \\ 
    1761.3719 & 18 & 210 &  8 &  8 & 1.42e+08 & 6.60e-02 && 4.39e+08 & \textbf{2.04e-01*} & \textbf{3.09e+00} && 5.71e+06 & 2.66e-03 & \textbf{4.03e-02} &&          &          & \textbf{        }   \\ 
    2088.2051 & 19 & 173 &  4 &  2 & 1.37e+08 & 4.48e-02 && 3.81e+08 & \textbf{1.25e-01*} & \textbf{2.79e+00} && 1.24e+08 & 4.05e-02 & \textbf{9.04e-01} &&          &          & \textbf{        }   \\ 
    2001.0284 & 21 & 190 & 12 & 10 & 9.53e+07 & 4.77e-02 && 2.27e+08 & \textbf{1.14e-01*} & \textbf{2.39e+00} && 7.26e+06 & 3.63e-03 & \textbf{7.61e-02} &&          &          & \textbf{        }   \\ 
    1798.1573 & 22 & 225 &  6 &  4 & 1.03e+08 & 3.33e-02 && 4.58e+08 & \textbf{1.48e-01*} & \textbf{4.44e+00} && 5.92e+06 & 1.91e-03 & \textbf{5.74e-02} &&          &          & \textbf{        }   \\ 
    1888.7343 & 23 & 218 & 10 & 10 & 1.42e+08 & 7.59e-02 && 3.24e+08 & \textbf{1.73e-01*} & \textbf{2.28e+00} && 7.43e+07 & 3.97e-02 & \textbf{5.23e-01} &&          &          & \textbf{        }   \\ 
    2527.0547 & 24 & 118 &  6 &  6 & 2.47e+08 & 2.36e-01 && 2.28e+08 & 2.18e-01 & \textbf{9.24e-01} && 1.52e+05 & 1.46e-04 & \textbf{6.19e-04} && 1.54e+08 & \textbf{1.48e-01*} & \textbf{6.27e-01}   \\ 
    2526.1482 & 25 & 123 & 14 & 14 & 1.91e+08 & 1.83e-01 && 2.61e+08 & 2.50e-01 & \textbf{1.37e+00} && 2.72e+08 & \textbf{2.60e-01*} & \textbf{1.42e+00} && 2.44e+08 & 2.34e-01 & \textbf{1.28e+00}   \\ 
    1932.4851 & 26 & 208 &  4 &  6 & 2.98e+07 & 2.50e-02 && 1.33e+08 & \textbf{1.12e-01*} & \textbf{4.48e+00} && 5.52e+07 & 4.64e-02 & \textbf{1.86e+00} &&          &          & \textbf{        }   \\ 
    2276.1804 & 27 & 166 & 12 & 12 & 1.08e+05 & 8.39e-05 && 2.42e+08 & \textbf{1.88e-01*} & \textbf{2.24e+03} && 1.81e+06 & 1.41e-03 & \textbf{1.68e+01} &&          &          & \textbf{        }   \\ 
    2499.6516 & 28 & 135 & 10 & 12 & 2.12e+08 & 2.38e-01 && 3.18e+06 & 3.57e-03 & \textbf{1.50e-02} && 2.95e+08 & 3.32e-01 & \textbf{1.39e+00} && 2.88e+08 & 3.24e-01 & \textbf{1.36e+00}   \\ 
    2535.1809 & 29 & 130 &  8 &  8 & 1.83e+08 & 1.76e-01 && 1.82e+08 & 1.75e-01 & \textbf{9.94e-01} && 8.34e+07 & \textbf{8.04e-02*} & \textbf{4.57e-01} && 2.35e+08 & 2.27e-01 & \textbf{1.29e+00}   \\ 
    2530.3191 & 30 & 131 &  4 &  4 & 2.02e+08 & 1.94e-01 && 1.21e+08 & 1.16e-01 & \textbf{5.98e-01} && 3.47e+05 & 3.33e-04 & \textbf{1.72e-03} && 2.93e+07 & \textbf{2.81e-02*} & \textbf{1.45e-01}   \\ 
    2569.1809 & 31 & 131 &  2 &  4 & 4.77e+07 & 9.44e-02 && 7.68e+07 & 1.52e-01 & \textbf{1.61e+00} && 3.26e+05 & 6.45e-04 & \textbf{6.83e-03} && 8.74e+07 & \textbf{1.73e-01*} & \textbf{1.83e+00}   \\ 
    2132.5479 & 31 & 186 &  2 &  4 & 1.31e+04 & 1.79e-05 && 1.11e+05 & 1.51e-04 & \textbf{8.44e+00} && 2.10e+08 & \textbf{2.86e-01*} & \textbf{1.60e+04} &&          &          & \textbf{        }   \\ 
    2480.9075 & 32 & 151 & 10 &  8 & 1.48e+08 & 1.09e-01 && 1.78e+07 & 1.31e-02 & \textbf{1.20e-01} && 2.08e+08 & 1.54e-01 & \textbf{1.41e+00} && 2.49e+08 & 1.84e-01 & \textbf{1.69e+00}   \\ 
    2424.8826 & 32 & 156 & 10 & 12 & 2.20e+08 & 2.33e-01 && 9.45e+07 & 1.00e-01 & \textbf{4.29e-01} && 1.03e+04 & 1.09e-05 & \textbf{4.68e-05} && 2.25e+08 & 2.38e-01 & \textbf{1.02e+00}   \\ 
    2547.4355 & 33 & 139 &  8 &  8 & 7.82e+07 & 7.61e-02 && 1.23e+08 & 1.20e-01 & \textbf{1.58e+00} && 9.37e+07 & 9.12e-02 & \textbf{1.20e+00} && 1.81e+08 & \textbf{1.76e-01*} & \textbf{2.31e+00}   \\ 
    2471.4170 & 33 & 153 &  8 &  6 & 1.54e+08 & 1.06e-01 && 1.80e+07 & 1.24e-02 & \textbf{1.17e-01} && 2.11e+08 & \textbf{1.45e-01*} & \textbf{1.37e+00} && 2.30e+08 & 1.58e-01 & \textbf{1.49e+00}   \\ 
    2630.3728 & 34 & 125 &  6 &  8 & 7.23e+07 & 1.00e-01 && 1.50e+06 & 2.07e-03 & \textbf{2.07e-02} && 7.04e+05 & 9.74e-04 & \textbf{9.74e-03} && 4.90e+07 & 6.78e-02 & \textbf{6.78e-01}   \\ 
    2467.5669 & 34 & 154 &  6 &  4 & 1.77e+08 & 1.08e-01 && 5.25e+07 & \textbf{3.19e-02*} & \textbf{2.95e-01} && 2.21e+08 & 1.34e-01 & \textbf{1.24e+00} && 2.28e+08 & 1.39e-01 & \textbf{1.29e+00}   \\ 
    2467.4185 & 35 & 155 &  4 &  2 & 2.64e+08 & 1.20e-01 && 9.19e+07 & \textbf{4.19e-02*} & \textbf{3.49e-01} && 2.91e+08 & 1.33e-01 & \textbf{1.11e+00} && 2.83e+08 & 1.29e-01 & \textbf{1.07e+00}   \\ 
     2464.0273 & 37 & 171 & 12 & 10 & 7.08e+07 & 5.37e-02 && 2.12e+08 & 1.61e-01 & \textbf{3.00e+00} && 2.04e+08 & 1.55e-01 & \textbf{2.89e+00} && 2.39e+08 & \textbf{1.81e-01*} & \textbf{3.37e+00}   \\ 
    2832.3945 & 38 & 129 &  4 &  6 & 7.65e+07 & 1.38e-01 && 1.08e+08 & 1.95e-01 & \textbf{1.41e+00} && 1.57e+08 & \textbf{2.83e-01*} & \textbf{2.05e+00} &&          &          & \textbf{        }   \\ 
    2506.8494 & 39 & 169 & 10 & 10 & 9.91e+07 & 9.34e-02 && 1.76e+08 & 1.66e-01 & \textbf{1.78e+00} && 3.32e+04 & 3.13e-05 & \textbf{3.35e-04} && 2.32e+08 & \textbf{2.18e-01*} & \textbf{2.33e+00}   \\ 
    2503.1475 & 40 & 170 &  8 &  8 & 1.43e+08 & 1.34e-01 && 1.54e+08 & 1.45e-01 & \textbf{1.08e+00} && 1.78e+06 & 1.67e-03 & \textbf{1.25e-02} && 2.23e+08 & \textbf{2.10e-01*} & \textbf{1.57e+00}   \\ 
    2462.6067 & 40 & 178 &  8 & 10 & 2.43e+08 & 2.76e-01 && 1.60e+08 & \textbf{1.82e-01*} & \textbf{6.59e-01} && 2.95e+07 & 2.68e-02 & \textbf{9.71e-02} && 2.60e+08 & 2.96e-01 & \textbf{1.07e+00}   \\ 
    2462.0288 & 41 & 180 &  6 &  8 & 2.34e+08 & 2.84e-01 && 1.65e+08 & \textbf{2.00e-01*} & \textbf{7.04e-01} &&          &          & \textbf{        } && 2.62e+08 & 3.17e-01 & \textbf{1.12e+00}   \\ 
    2768.3208 & 42 & 146 & 12 & 14 & 1.58e+08 & 2.12e-01 && 2.03e+06 & 2.72e-03 & \textbf{1.28e-02} && 1.93e+08 & \textbf{2.59e-01*} & \textbf{1.22e+00} &&          &          & \textbf{        }   \\ 
    2841.4861 & 44 & 141 &  2 &  4 & 7.62e+07 & 1.84e-01 && 8.34e+07 & 2.02e-01 & \textbf{1.10e+00} && 8.56e+07 & \textbf{2.07e-01*} & \textbf{1.12e+00} &&          &          & \textbf{        }   \\ 
    2775.5066 & 44 & 152 &  2 &  4 & 2.73e+07 & 6.31e-02 && 1.88e+07 & 4.34e-02 & \textbf{6.88e-01} && 5.38e+07 & \textbf{1.24e-01*} & \textbf{1.97e+00} &&          &          & \textbf{        }   \\ 
    2543.4995 & 44 & 173 &  2 &  2 & 1.61e+08 & 1.56e-01 && 9.47e+07 & 9.18e-02 & \textbf{5.88e-01} && 2.10e+08 & \textbf{2.04e-01*} & \textbf{1.31e+00} &&          &          & \textbf{        }   \\ 
    2519.8059 & 45 & 181 &  8 &  6 & 2.10e+08 & 1.50e-01 && 1.93e+08 & 1.38e-01 & \textbf{9.20e-01} && 2.54e+08 & \textbf{1.81e-01*} & \textbf{1.21e+00} &&          &          & \textbf{        }   \\ 
    2450.9475 & 50 & 197 &  2 &  4 & 1.26e+08 & 2.27e-01 && 1.65e+08 & 2.97e-01 & \textbf{1.31e+00} && 1.03e+08 & 9.28e-02 & \textbf{4.09e-01} && 1.86e+08 & \textbf{3.36e-01*} & \textbf{1.48e+00}   \\ 
    2448.3083 & 50 & 198 &  2 &  2 & 7.45e+06 & 6.69e-03 && 5.90e+05 & 5.30e-04 & \textbf{7.92e-02} && 1.16e+08 & 5.21e-02 & \textbf{7.79e+00} && 1.52e+08 & \textbf{1.36e-01*} & \textbf{2.03e+01}   \\ 
    2441.1633 & 51 & 201 &  6 &  8 & 1.18e+08 & 1.41e-01 && 1.65e+08 & 1.97e-01 & \textbf{1.40e+00} && 2.06e+08 & 2.45e-01 & \textbf{1.74e+00} && 2.31e+08 & \textbf{2.75e-01*} & \textbf{1.95e+00}   \\ 
    2425.3276 & 51 & 204 &  6 &  6 & 1.24e+08 & 1.09e-01 && 1.41e+08 & 1.24e-01 & \textbf{1.14e+00} && 1.45e+08 & 1.28e-01 & \textbf{1.17e+00} && 1.50e+08 & \textbf{1.32e-01*} & \textbf{1.21e+00}   \\ 
    \hline    
  \end{tabular}}
\end{table*}



\bsp	
\label{lastpage} 
\end{document}